\documentclass[]{aastex631}
\usepackage{microtype}
\usepackage{multirow}
\usepackage{gensymb}
\usepackage{xcolor}
\usepackage{amsmath}
\usepackage[bottom]{footmisc}
\usepackage{comment}
\usepackage{float}
\usepackage{xcolor}

\newcommand{\msol}{M_{\odot}}
\newcommand{\rsol}{R_{\odot}}
\newcommand{\mwd}{M_{\rm WD}}
\newcommand{\rwd}{R_{\rm WD}}
\newcommand{\astar}{a^{\star}}

\newcommand{\tb}{\textcolor{black}}
\begin{document}
\title{Tidal encounters of close white dwarf binaries with spinning black holes}
\correspondingauthor{Aryabrat Mahapatra}
\email{aryabratm23@iitk.ac.in}

\author{Aryabrat Mahapatra}
\affiliation{Department of Physics, Indian Institute of Technology Kanpur,
	Kanpur 208016, India}

\author{Adarsh Pandey}
\affiliation{Department of Physics, Indian Institute of Technology Kanpur,
	Kanpur 208016, India}

\author{Pritam Banerjee}
\affiliation{Indian Institute of Astrophysics, Koramangala, Bangalore 560034, India}

\author{Tapobrata Sarkar}
\affiliation{Department of Physics, Indian Institute of Technology Kanpur,
	Kanpur 208016, India}

\begin{abstract}
When a stellar binary encounters a spinning black hole, interesting phenomena might result due to
the mutual interaction between the binary spin, orbital angular momentum and the black hole spin. 
Here we consider such encounters between an intermediate mass spinning black hole and a close identical white 
dwarf binary system whose center of mass follows a parabolic trajectory. After studying a corresponding 
three-body problem in the point particle approximation, we perform 
a suite of smoothed particle hydrodynamics based numerical simulations of such scenarios. For this, we integrate the 
geodesic equations for the spinning black hole, while considering the hydrodynamics and the self and 
mutual gravitational interactions of the stars in a Newtonian approximation, an approach justified by the 
choice of parameters in the theory. We consider various
initial configurations of the binary center of mass leading to equatorial and off-equatorial orbits, as also various 
initial inclinations between the binary's initial spin angular momentum and its initial orbital angular momentum. 
We find that the effects of black hole spin manifest in the tidal dynamics of the binary components, while 
the observables of tidal encounters such as mass fallback rates are strongly dependent on 
the initial inclination angle.
We show that the influence of the black hole spin emerges in distinct ways for different initial configurations of the binary’s spin alignment. We establish that
within the ambits of the Hills mechanism, in certain cases, the fallback rate may show a
three-hump structure, due to interactions between tidal debris of the individual stars. 
\end{abstract}
	
\section{Introduction}
\label{sec1}

A binary stellar system that encounters a black hole is a classic example of a three body problem. Such problems
have been studied for more than three centuries now, and a modern exposition can be found in the textbook of \cite{VK}. 
The problem assumes importance in the context of astrophysical phenomena, as a significant number of
stars are found in binary configurations \citep{Toonen}. When a binary stellar system comes close enough to a black hole,
tidal disruption events (TDEs) often result in observable signatures, that provides an ideal laboratory to test existing theories 
of gravitational interactions. In the context of three body problems involving a black hole and a stellar binary, point particle 
approximations for stars are easier to handle analytically and numerically, since one can integrate the geodesic equations 
under certain approximations of the self gravity between the two stars. However, the situation becomes 
more complicated when one considers stellar hydrodynamics as well. This is because stellar deformations and possible disruptions 
non-trivially affect the outcome of close encounters of binary stellar systems with black holes. Then, in the absence of
enough symmetries, analytical computations are somewhat challenged. Numerical simulations are often employed to determine
the nature of interaction in these cases, and a popular method is that of smoothed particle hydrodynamics (SPH). 

\tb{In a recent previous paper \citep{binary1}, we studied tidal encounters of close non-accreting white dwarf (WD)
	binaries with a non-rotating intermediate-mass black hole (IMBH). We briefly summarize below the astrophysical context motivating such interactions before describing the present extension to rotating Kerr IMBHs.}

\tb{Encounters between close WD binaries and much more massive black holes are expected to occur naturally in dense stellar environments. In the case of IMBHs, globular clusters provide one of the most plausible astrophysical settings. 	Recent observations of the globular cluster $\omega$ Centauri provide strong
	evidence for an IMBH in such an environment. 
\cite{Haeberle2024} identified seven fast-moving stars within
	the central $3$ arcsec ($\simeq0.08$ pc) of $\omega$ Centauri, consistent with
	the presence of a BH of mass $\gtrsim 8\times10^3\,M_\odot$. These systems are among the most promising candidates for hosting IMBHs, where high stellar densities promote runaway stellar collisions, repeated mergers of stellar-mass black holes, and efficient dynamical interactions \citep{IMBH0, 2008ApJ...676.1008N, 2002ApJ...578L..41G}. At the same time, close double white dwarf (DWD) binaries are expected to be abundant in such environments because binary evolution and frequent dynamical encounters efficiently produce and harden compact binaries \citep{shara, Ivanova, Ivanova2}. Mass segregation and two-body relaxation gradually drive these compact binaries toward the cluster core, where close encounters with the central IMBH become increasingly likely. Numerical studies suggest stellar tidal disruption rates of $\sim10^{-7}$--$10^{-6} ~\mathrm{yr^{-1}}$ per globular cluster containing an IMBH \citep{Baumgardt2004}, while population synthesis calculations predict several DWD systems can be produced per Gyr in typical massive clusters \citep{Ivanova2}. Consequently, although DWD tidal disruption events are expected to be intrinsically rare, globular clusters provide a physically well-motivated environment in which compact WD binaries may be scattered onto loss-cone orbits and undergo tidal separation or disruption by an IMBH. Compact binaries in galactic nuclei hosting massive black holes have also attracted considerable attention, where relaxation processes, binary-single encounters, and secular evolution (e.g., eccentric Kozai-Lidov oscillations) can significantly modify binary evolution, producing mergers, tidal separations, or tidal disruption events \citep{Stephan2016, Stephan2019, Zhu2018, Wang2021b}. Thus, interactions between compact binaries and central black holes constitute a broad astrophysical problem extending beyond the classical Hills mechanism. The present study focuses on the dynamics of a single close encounter between a WD binary and an IMBH. Our objective is to investigate the hydrodynamical consequences of an individual close passage, rather than the long-term dynamical evolution of binaries in galactic nuclei.}	

Here, we extend this analysis 
to rotating (Kerr) IMBHs. There are several motivations for the analysis performed in this paper. For example, most observed black
holes are known to be spinning, with recent evidence suggesting that many of these can have spin parameters close to
extremality. Thus studying the phenomenology associated with spinning BHs is important and interesting in its own right, more
so for IMBHs, whose observational evidence has been relatively rare compared to their super massive BH cousins. 
Stellar objects that have intrinsic spin are of further interest here, as their ``coupling'' with BH spin often results in 
rich physics, as was detailed in \cite{Garain2} for tidal encounters of spinning WDs with Kerr BHs. Binary star systems
with large intrinsic angular momenta are expected to capture this feature to a greater extent. Also note that for spinning BHs
which possess axial symmetry, off-equatorial orbits are important and interesting to study and a significant amount of
literature exists on tidal interactions of solitary stars in such orbits. Stellar binaries assume importance in this context,
since there exists larger freedom for the initial configuration of the system, which includes for example the relative 
interplay of the initial binary angular momentum about its centre of mass with the orbital angular momentum and the
inclination of the orbit. All these put together makes the study of binary stellar systems in spinning black hole backgrounds
important and interesting. 

Here we present the results of a suite of SPH simulations that capture the effects of a spinning black hole background on
a close WD binary. Our simulations integrate the Kerr BH geodesics along with a Newtonian treatment of the hydrodynamics
of individual star, as well as the self gravity between them, 
as detailed in \cite{binary1}. To justify this approach, we note that the pericenter distance in our analysis
is fixed at $25 r_g$, with $r_g = GM/c^2$ being the gravitational radius, where Newtonian hydrodynamics remains a 
sufficiently robust approximation, see \cite{Stone2019}, \cite{2017MNRAS.469.4483T}. Even at this ``large'' pericenter
distance, the effects of the BH spin indicate richer physics compared to the non-spinning black hole, as we will
argue in sequel. 
To the best of our knowledge, this scenario has not been studied in the literature, although studies on TDEs of 
solitary stars in equatorial and off-equatorial orbits in Kerr BH backgrounds are fairly abundant. For a sampling of the literature, see  
the works of \cite{RR, RR1, Beloborodov, WigginsLie, Ivanov, Ishii, 
Ferrari, Kesden1, SinghKesden, Haas, Hayasaki2016, 2017MNRAS.469.4483T, Liptai2019, Banerjee, GaftonRosswog, Jankovic}. 
Further details and a more exhaustive set of references appear in \cite{binary1}. 

\section{Methodology and Initial Setup}
\label{sec2}

In this section, we provide a brief description of the numerical methodology employed to simulate the tidal disruption of close WD binaries in a generic 
off-equatorial parabolic orbit around a spinning IMBH. 
\cite{Banerjee2023} presents all details on our SPH code to simulate the tidal disruption of a stellar fluid in the external gravitational field of a BH,
that is inspired by the publicly available code PHANTOM \citep{2018PASA}. To efficiently compute fluid properties and inter-particle forces, we employ a binary tree structure with an opening angle parameter set to $\theta_{\rm MAC} = 0.5$. We adopt the standard SPH artificial-viscosity parameters $\alpha^{\rm AV} = 1.0$ and $\beta^{\rm AV} = 2.0$ \citep{Hayasaki2013}, along with the Balsara switch to suppress viscosity in shear flows \citep{Balsara1995}.
The SPH fluid equations are integrated using a leapfrog integrator (with global time-stepping), ensuring the exact conservation of energy and angular momentum throughout the simulation. To achieve a single stable WD configuration in SPH, we adopt the methodology outlined in \cite{Garain2023}. We obtain the output density profile representing a carbon-oxygen WD in hydrostatic equilibrium from the stretch-mapping process \citep{1994MmSAI..65.1013H} in SPH. The final relaxed state of the WD used here satisfies the virial equilibrium condition, with deviations from viriality below $1\%$. Next, we introduce a close, detached circular binary of two identical low-mass WDs following the setup described in \cite{binary1}. Circular binaries are a reasonable initial choice, as close binaries in nature tend to circularize more rapidly than wide binaries, and ultimately settle into nearly circular orbits. Nevertheless, the eccentricity of the binary system can evolve significantly in the BH background, changing repeatedly as the system undergoes tidal separation during its interaction with the BH. We ensure that individual WDs experience no tidal deformations due to their companion, and that orbital oscillations remains free from dissipations or drift. The binary system is evolved for $\geq 15$ orbital cycles, with deviations kept below $1\%$. Finally, the stable WD binary system is placed in the background of a rapidly spinning (Kerr) IMBH.

To investigate the tidal interactions with the centre of mass of the binary in a generic parabolic orbit, we adopt a hybrid formalism, in which we integrate the Kerr geodesics while implementing a Newtonian form of the fluid self-gravity, as we have mentioned before. The Kerr space-time in Boyer–Lindquist (BL) coordinates $(t,r,\theta,\phi)$ is given by
\begin{eqnarray}
	ds^2 &=& -\left(1-\frac{2GMr}{c^2\Sigma}\right)c^2dt^2 - \frac{4GMra\sin^2{\theta}}{c\Sigma}dtd\phi + \frac{\Sigma}{\Delta}dr^2 
	+ \Sigma d\theta^2  +  \left(r^2+a^2+\frac{2GMra^2\sin^2{\theta}}{c^2\Sigma}\right)\sin^2{\theta}d\phi^2 
	\label{KerrMetric}
\end{eqnarray}
where $\Sigma = r^2 + a^2\cos^2\theta$ and $\Delta = r^2 - 2GMr/c^2 + a^2$, and $c$ denotes the speed of light.
The Kerr spin parameter is $a = J/(Mc)$, with $M$ and $J$ denoting the BH mass and angular momentum. The dimensionless spin parameter is defined as $a^\star = a/r_g$. To set the WD binary on an off-equatorial parabolic orbit, we must specify appropriate initial conditions: the initial coordinates and the initial velocity of the binary’s centre of mass (CM), denoted $(r_0, \theta_0, \phi_0)$ and $(\dot{r}_0, \dot{\theta}_0, \dot{\phi}_0)$, respectively. These are obtained by determining the turning points through the root finding approach of $dr/d\tau$, as done in \cite{Garain3}, following the procedure mentioned in \cite{2017MNRAS.469.4483T}. For a parabolic trajectory, we require to solve the third-order polynomial equation 
\begin{eqnarray}
	\mathcal{R}(r) = c^4 r_s r^3 - c^2 (l_z^2 + \mathcal{Q})r^2 + l^2 r_s r -a^2 c^2 \mathcal{Q}~,
\end{eqnarray}
where $r_s = 2r_g$, $l^2$ is the square of total magnitude of specific angular momentum and $\mathcal{Q}$ is the Carter constant. The polynomial corresponds to the roots $r_p$, $r_c$ and $r_d$. The pericentre distance $r_p$ is evaluated from the definition of binary's impact parameter; $\beta^b=r_t^b/r_p$, where tidal radius of binary is given by:  
\begin{eqnarray}\label{r_t}
r_t^b = R_{\rm sep}\left(\frac{M}{M_b}\right)^{1/3}
\end{eqnarray}
 Here, we consider $R_{\rm sep}=a_0/2$, where, $a_0$ is the initial separation between the two stars in binary and $M_b$ is the total mass of the binary system. The constants of motion can be written in terms of these roots as
\begin{equation}\label{eq4}
	\begin{aligned}
		\mathcal{Q} = \frac{c^2 r_s r_p r_c r_d}{a^2}~,~~
		l^2 = c^4(r_p (r_c + r_d) + r_c r_d)~,~~ l_z^2 = c^2 r_s\frac{a^2(r_p + r_c + r_d)-r_p r_c r_d}{a^2}~.
	\end{aligned}
\end{equation}
Additionally, we use the relation $l^2 = c^2\mathcal{Q} + (l_z c-a \mathcal{E})^2$ and for a desired off-equatorial orbit with initial polar angle, 
$\theta_0=\theta_a$ (to avoid any visual conflict with $\vartheta_0$ introduced later), we have $d\theta/d\tau = 0 \implies \mathcal{Q}=l_z^2\cot^2{\theta_a}$.
\tb{Here, the initial polar angle $\theta_a$ is measured from the $+z$-axis, which is aligned (or anti-aligned) with the BH spin axis, to the initial radial direction ($\hat{r}_0$) of the binary’s CM.}
 Once the roots are solved and the constants of motion are evaluated from Equation \ref{eq4}, then, substituting these into the first integrals of Kerr geodesics yield the initial velocity of the binary’s CM.
\begin{figure}[h]
	\epsscale{0.8}
	\plotone{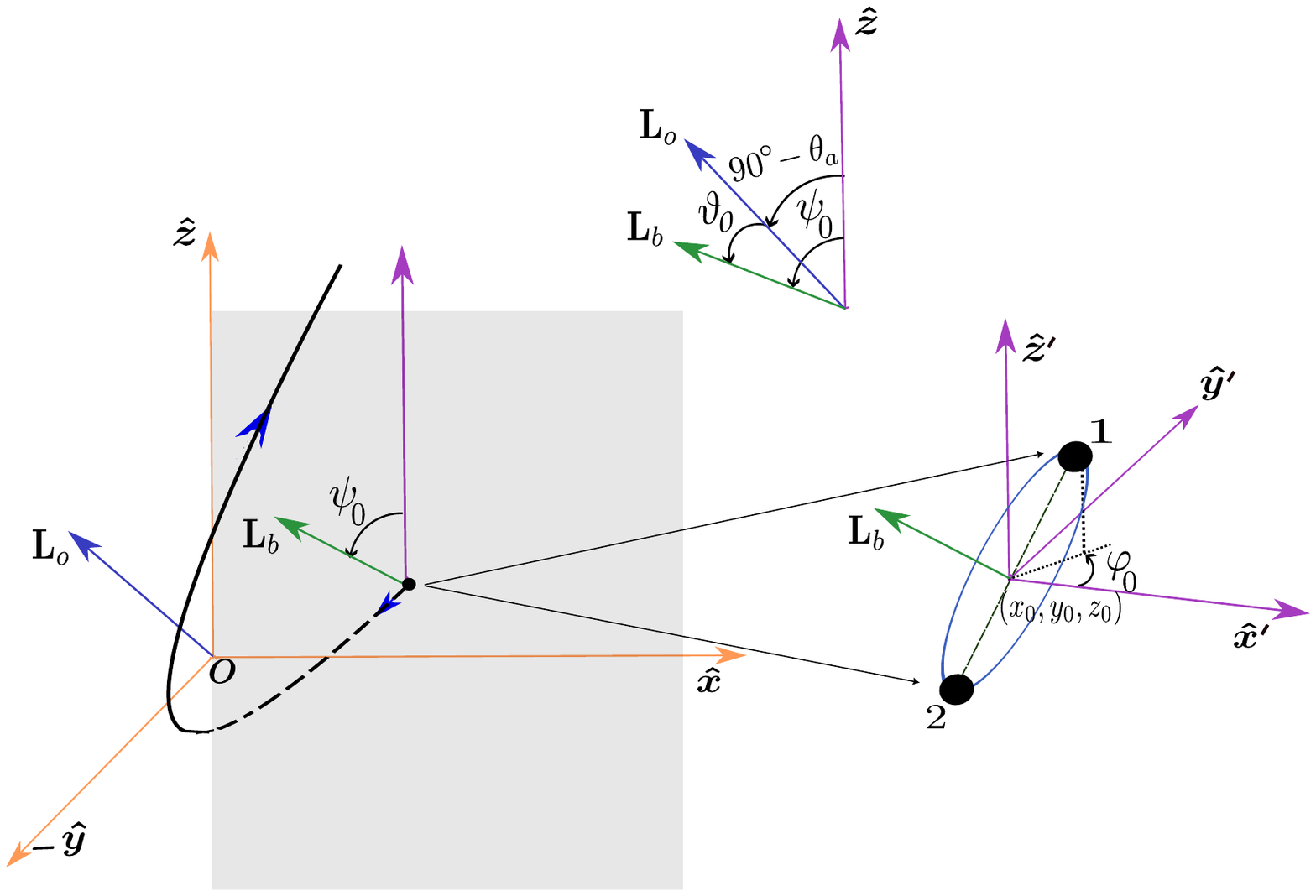}
	\caption{\small Qualitative depiction of binary stellar dynamics around a spinning BH.}
	\label{img}
\end{figure}

A qualitative depiction of a generic binary system on a parabolic orbit around a spinning BH is given in Figure \ref{img}. 
The BH located at the origin $O$, with its spin either parallel or antiparallel to the $z$ axis. The centre of mass of the binary is initialized at $(x_0, y_0, z_0)$, while its two components (which we take to be identical WDs) have their CMs at $(x_{10}, y_{10}, z_{10})$ and $(x_{20}, y_{20}, z_{20})$, separated by a distance $a_0$. The initial angular momentum of the binary about its centre of mass ($\mathbf{L}_b$) is oriented at an angle 
$\vartheta_0$ relative to the initial orbital angular momentum (OAM) ($\mathbf{L}_o$) of the binary around the BH, computed about the origin. The binary phase angle, $\varphi_0$ denotes the initial angle made by the line joining the binary components with the $\hat{x}$ direction. By convention, angles are measured as positive in the directions indicated in the figure.

We choose the initial CM coordinates at a large distance from the BH, $r_0 = 10 r_t^b \simeq 500 r_g$. Further, to simplify the
analysis, coordinates are chosen such that $\phi_0=0$ (so that the CM is initially on the \(x\text{--}z\) plane). 
The initial Cartesian position of the binary CM,
\(\mathbf{r}_{\mathrm{CM}} = (x_0, y_0, z_0)\), is obtained via the standard
transformation from spatial BL to Cartesian coordinates. The
corresponding initial CM velocity,
\(\mathbf{v}_{\mathrm{CM}} = (v_{x0}, v_{y0}, v_{z0})\), is computed by the similar transformation from the BL coordinate velocities
\((\dot{r}_0, \dot{\theta}_0, \dot{\phi}_0)\).

According to our convention, a circular WD binary with initial separation
\(a_0\), binary inclination \(\vartheta_0\), and initial phase \(\varphi_0\)
(see Figure~\ref{img}) is constructed by applying two successive Euclidean
rotations. This procedure provides an excellent approximation at the large
initial separations from the BH considered here, where spacetime curvature
effects across the binary scale are negligible. Initially, we adopt a primed
Cartesian coordinate system \((x', y', z')\) that coincides with the global
Cartesian axes \((x, y, z)\), and place the binary CM at the origin.

We begin by positioning the binary along the \(x'\)-axis. An anticlockwise rotation about the
\(z'\)-axis by an angle \(\varphi_0\) sets the initial orbital phase of the
binary. This is followed by a clockwise rotation about the \(+y'\)-axis by an
angle \(\psi_0\), which imposes the desired binary inclination
\(\vartheta_0\). Finally, the binary CM is translated to its initial global
position \(\mathbf{r}_{\mathrm{CM}}\), with the BH fixed at the origin
of the coordinate system.

The initial Cartesian positions of the individual WD centers of mass,
\(\mathbf{x}_{i0} = (x_{i0}, y_{i0}, z_{i0})\), are then given by
\begin{eqnarray}
	\mathbf{x}_{i0} = \mathbf{r}_{\mathrm{CM}} \pm \frac{a_0}{2}
	\Big(\cos\psi_0 \cos\varphi_0,\,
	\sin\varphi_0,\,
	\sin\psi_0 \cos\varphi_0\Big),
\end{eqnarray}
where the plus (minus) sign corresponds to WD1 (WD2).

The corresponding initial velocities are
\begin{eqnarray}
	\mathbf{v}_{i0} = \mathbf{v}_{\mathrm{CM}} \pm v_b
	\Big(-\cos\psi_0 \sin\varphi_0,\,
	\cos\varphi_0,\,
	-\sin\psi_0 \sin\varphi_0\Big),
\end{eqnarray}
where $\psi_0 = (90^\circ - \theta_a + \vartheta_0)$, and $v_b = \sqrt{G M_{\mathrm{WD}}/2 a_0}$
denotes the characteristic orbital velocity of each WD within the binary.

The values for these initial conditions are finally determined by the tidal disruption parameters adopted in this study. We consider a tidal encounter between a rapidly spinning IMBH of mass $8\times10^{3}\,\msol$, motivated by recent observations of an IMBH candidate in $\omega$ Centauri \citep{Haeberle2024}, with spin parameter $a^{\star}=\pm 0.98$ and an identical WD binary, where each WD has mass $\mwd = 0.5\,\msol$, radius $\rwd = 0.0141\,\rsol$, and initial separation $a_0 = 6~\rwd$. \tb{For reference, the corresponding gravitational radius is $r_g=GM_{\rm BH}/c^2\simeq 0.0169~\rsol$, while the Kerr event-horizon radius is, $r_+=r_g[1+\sqrt{1-(a^\star)^2}]\simeq 0.0204~\rsol$.} To ensure that the interaction remains around a meaningful regime of our interest at $r_p \simeq 25~r_g$, we set $\beta^b = 2.0$, following from Equation \ref{r_t}. 

Finally, we assign the remaining parameter values in order to explore how the tidal disruption dynamics depend 
on $a^\star$, $\mathbf{L}_b$ and $\mathbf{L}_o$. 
To this end, for our simulations, we consider either an off-equatorial orbit close to the meridional plane (defined by $\theta_a=0^\circ$)
with an initial value $\theta_a=1^\circ$ or an equatorial orbit with an initial value $\theta_a=90^\circ$. 
In the equatorial orbit, the binary has prograde motion around the BH when $a^\star=+0.98$ and retrograde motion when $a^\star=-0.98$. 
In these cases, we expect tidal effects to be maximal or minimal, as discussed in \cite{Garain3}. 
As there are a large number of parameters appearing in our formulation, we will simplify the analysis
by fixing the initial phase, $\varphi_0 = 0^\circ$, and alter the inclination angle $\vartheta_0$. The essential physics of 
the phenomena that we study are captured in this choice, although the initial phase might play an significant role in 
determining some of the details, as pointed out in \cite{binary1}. In this work our intention is to understand the variation of the angle $\theta_a$ along with its offset $\vartheta_0$ to exploit the axisymmetry of a Kerr black hole. We understand that the combinations of all three angles ($\theta_a$, $\vartheta_0$, $\varphi_0$) is the ideal choice of study. However, due to its large parameter space we choose not to concentrate on the azimuthal offset phase in this paper. Rather, a culmination of all parameters is kept as a future work. We will make a few comments on the choice of the initial phase in
the following Section \ref{spin_coupling} and in the concluding section of this paper.

Throughout this work, we adopt the convention that \(\vartheta_0 = 0^\circ\) corresponds to prograde rotation of the binary spin relative to the OAM of the binary around the BH, while \(\vartheta_0 = 180^\circ\) corresponds to retrograde rotation\footnote{A retrograde configuration can be obtained either by reversing the initial binary spin or by rotating the binary by $180^\circ$; the two are equivalent for an identical binary. In our simulations, we adopt the former approach, as both choices yield identical dynamics up to a relabelling of the stars. The same equivalence applies to any inclination $\vartheta_0$ and its counterpart $\pi + \vartheta_0$.}. This nomenclature will be used consistently in what follows. We also consider configurations with \(\vartheta_0 = 90^\circ\), which represent cases where the binary spin lies on the orbital plane, directed towards the BH.
For off-equatorial orbits, when the binary spin is aligned with the \(+z\) axis (\(-z\) axis), the initial phase angle is set to \(\psi_0 = 0^\circ\) (\(\psi_0 = 180^\circ\)). These choices correspond to binary inclinations
\[
\vartheta_0 = \theta_a - 90^\circ \quad (\text{spin along } +z), \qquad
\vartheta_0 = \theta_a + 90^\circ \quad (\text{spin along } -z).
\]
For an equatorial orbit (\(\theta_a = 90^\circ\)), these relations reduce to \(\vartheta_0 = 0^\circ\) and \(\vartheta_0 = 180^\circ\), respectively. In contrast, for an off-equatorial orbit with \(\theta_a = 1^\circ\), they correspond to \(\vartheta_0 = -89^\circ\) and \(\vartheta_0 = 91^\circ\). Consequently, this makes our analysis restricted to configurations with specific initial binary inclinations $\vartheta_0 = 0^\circ$, $\vartheta_0 \simeq \pm 90^\circ$ and $\vartheta_0 = 180^\circ$.

We also briefly comment on the possible effects of tidal heating during the tidal compression and stretching of WDs in the gravitational field of the IMBH. In sufficiently deep encounters, extreme tidal compression may produce substantial heating in the WD and potentially trigger thermonuclear runaway events, often referred to as ‘tidal novae’. However, for the encounter strengths and parameter regime considered in the present work, such effects are expected to be negligible. This conclusion is supported by the detailed timescale and thermodynamic estimates presented in \cite{binary1}.

\section{Point-particle dynamics in the Three-Body Problem}\label{point_particle}

In this section, we describe the dynamics of the binary system by treating the stellar components as an initially bound pair of test particles orbiting a spinning BH. This simplified setup enables a comparative analysis for understanding how tidal interactions operate when the finite sizes of the stellar objects are later taken into account. Such finite-size effects can significantly modify the tidal dynamics through both the mutual gravitational influence of the companion and the hydrodynamics of each individual stellar component.

\subsection{Tidal separation and binary spin effects}\label{sec:binary_spin}

The evolution of a bound pair of particles constituting a circular binary and moving on a parabolic orbit around a spinning BH is governed by
\begin{eqnarray}\label{eom}
\ddot{x}^i_a = - \left( g^{i\lambda} - g^{0\lambda} \dot{x}^i \right)_a \left[ \left( \frac{\partial g_{\mu\lambda}}{\partial x^\nu} - \frac{1}{2} \frac{\partial g_{\mu\nu}}{\partial x^{\lambda}} \right) \dot{x}^{\mu} \dot{x}^{\nu} \right]_a - \frac{G m_b}{|x^i_a - x^i_b|^3} (x^i_a - x^i_b)~.
\end{eqnarray}

\begin{figure}[h]
	\centering
	\gridline{
		\fig{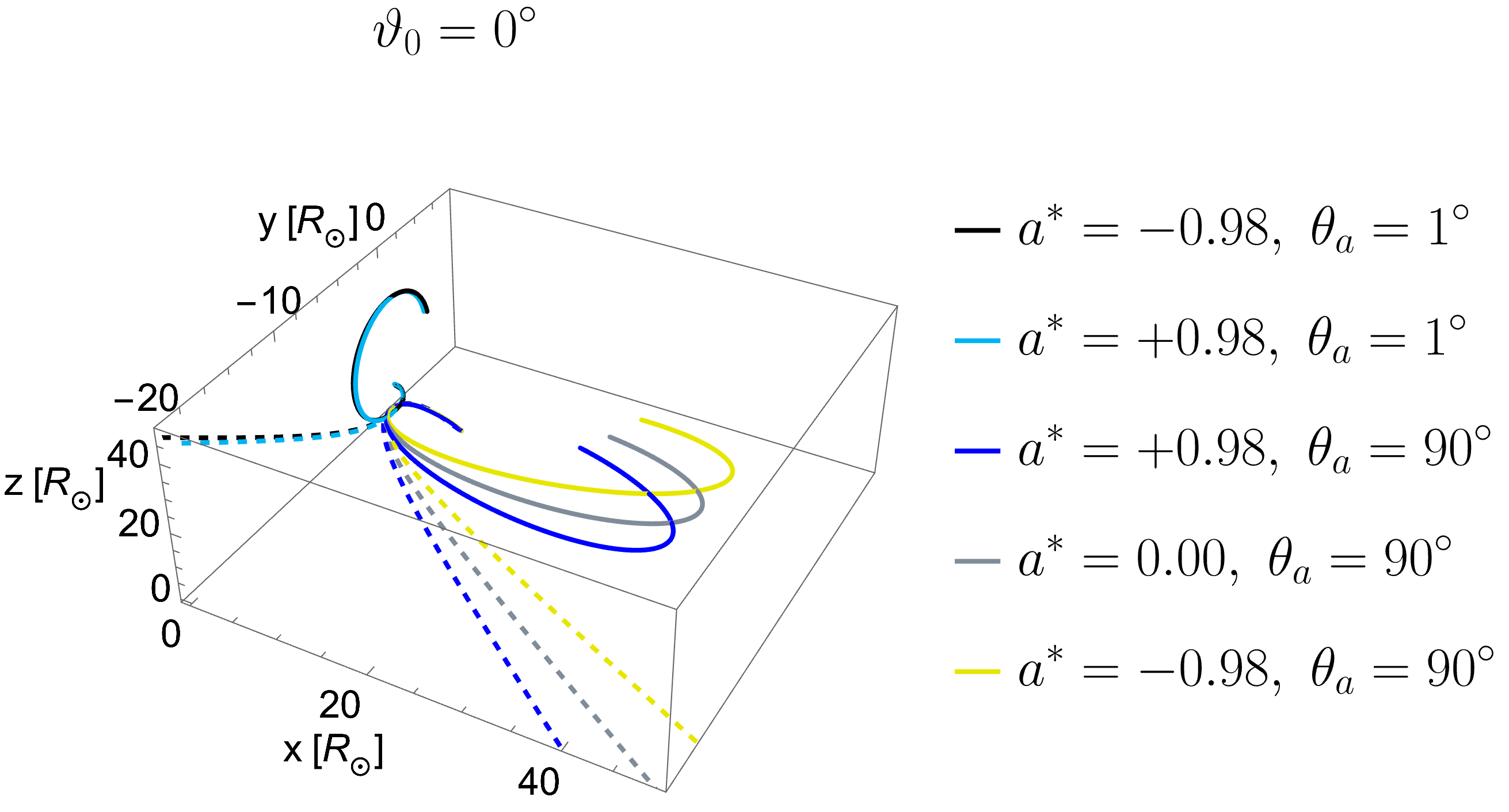}{0.58\textwidth}{}
		\fig{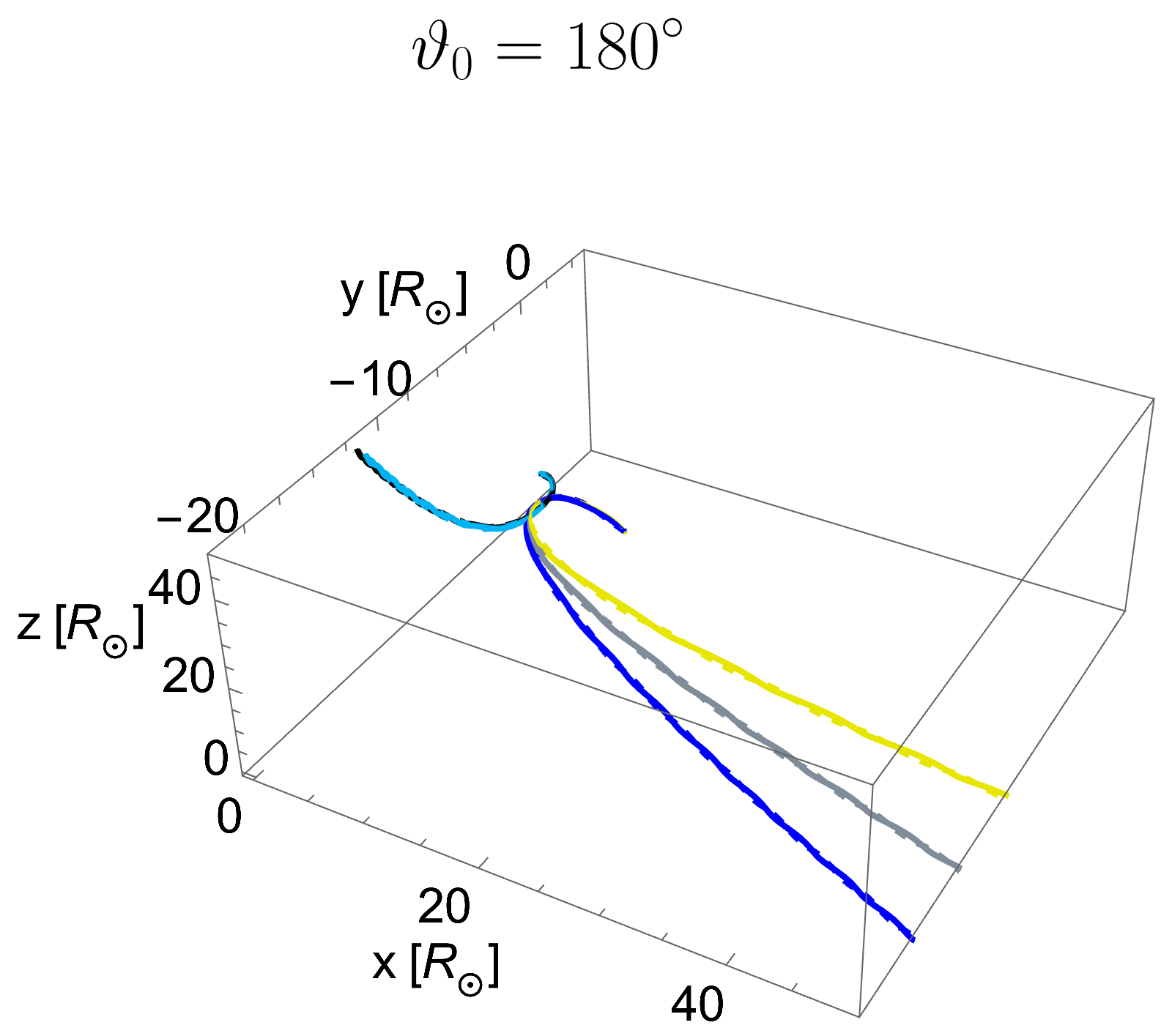}{0.35\textwidth}{}
	}
	\caption{{\small Trajectories of binary test particle around the spinning BH. \tb{The solid and dashed lines denote the point-masses $m_1$ and $m_2$ respectively. The trajectories are plotted for both spinning and non-spinning (Schwarzschild) BH. For spinning BH, equatorial and off-equatorial orbits are displayed corresponding to $\theta_a=90^\circ$ and $\theta_a=1^\circ$ respectively with BH spin parameter $\astar=\pm0.98$. The different colors correspond to the different BH spin parameters and initial polar angles indicated in the legend.} \textbf{Left Panel: }The initial motion of the binary is prograde relative to the OAM ($\vartheta_0=0^\circ$). \textbf{Right Panel: }The initial motion of the binary is retrograde relative to the OAM ($\vartheta_0=180^\circ$).}}
	\label{xyz1}
\end{figure}

Here, the first term describes the motion of particles in the Kerr spacetime background and the second term accounts for the mutual self-gravity of the two particles.  \tb{The second term can be treated to a very good approximation in the Newtonian limit. First, the size of the binary is sufficiently small compared with the characteristic curvature scale of the central BH spacetime at the site of encounter. For example, for an equatorial encounter around a spinning BH, we can estimate the local curvature by the following relative length scale:
\begin{equation}\label{alpha}
	\alpha = \frac{a_0}{\mathcal{R}}, ~~\text{where the local radius of curvature }~ \mathcal{R}=\left(\frac{r}{r_g}\right)^{3/2}r_g
\end{equation}
For our parameters, at pericentre this gives $\mathcal{R}=125~r_g$, yielding $\alpha\simeq0.04$. Therefore, when $\alpha\ll1$, relativistic corrections associated with the external curvature across the binary are expected to be small. In Appendix \ref{appendix1}, we further quantify the deviation introduced by relativistic corrections by incorporating the $1\mathtt{PN}$ correction into the self-gravity term of Equation \ref{eom}, including cases where the binary separation becomes larger.}
 Using the solutions of Equation \ref{eom}, we compute the individual particle trajectories for $\astar=\pm 0.98$ and $\theta_a=1^\circ,~90^\circ$. In the left panel of Figure \ref{xyz1}, we show the trajectory for a binary whose initial binary spin has prograde rotation $(\vartheta_0 = 0^\circ)$, while the corresponding retrograde case $(\vartheta_0 = 180^\circ)$ is presented in the right panel. 

We find that, for all combinations of parameters in the prograde rotation, the binary system undergoes tidal separation, one particle is captured onto a bound orbit around the BH, while the other is ejected onto a parabolic-like trajectory. This behaviour indicates the well-known Hills mechanism \citep{Hills}. In contrast, for retrograde rotation, the binary system preserves its bound state throughout the entire evolution, where the dynamics is not governed through Hills mechanism. For equatorial $\theta_a=90^\circ$ orbits, the stronger apsidal precession produced for retrograde motion, 
drags the binary particle trajectories significantly closer to the BH compared to the prograde case. However, it is almost same in $\theta_a=1^\circ$ orbit for both $\astar=\pm0.98$. Moreover, in both $\theta_a = 90^\circ$ and $\theta_a = 1^\circ$ configurations, the particle motion remains effectively confined to the orbital plane around the BH. Only a small deviation appears in the inner orbital motion of the binary for the $\theta_a = 1^\circ$ case. Note that by ``inner orbital motion'' we refer to the motion of the binary about its CM, whereas ``outer orbital motion'' indicates the motion of the binary around the BH. 

\begin{figure}[h]
	\epsscale{0.75}
	\plotone{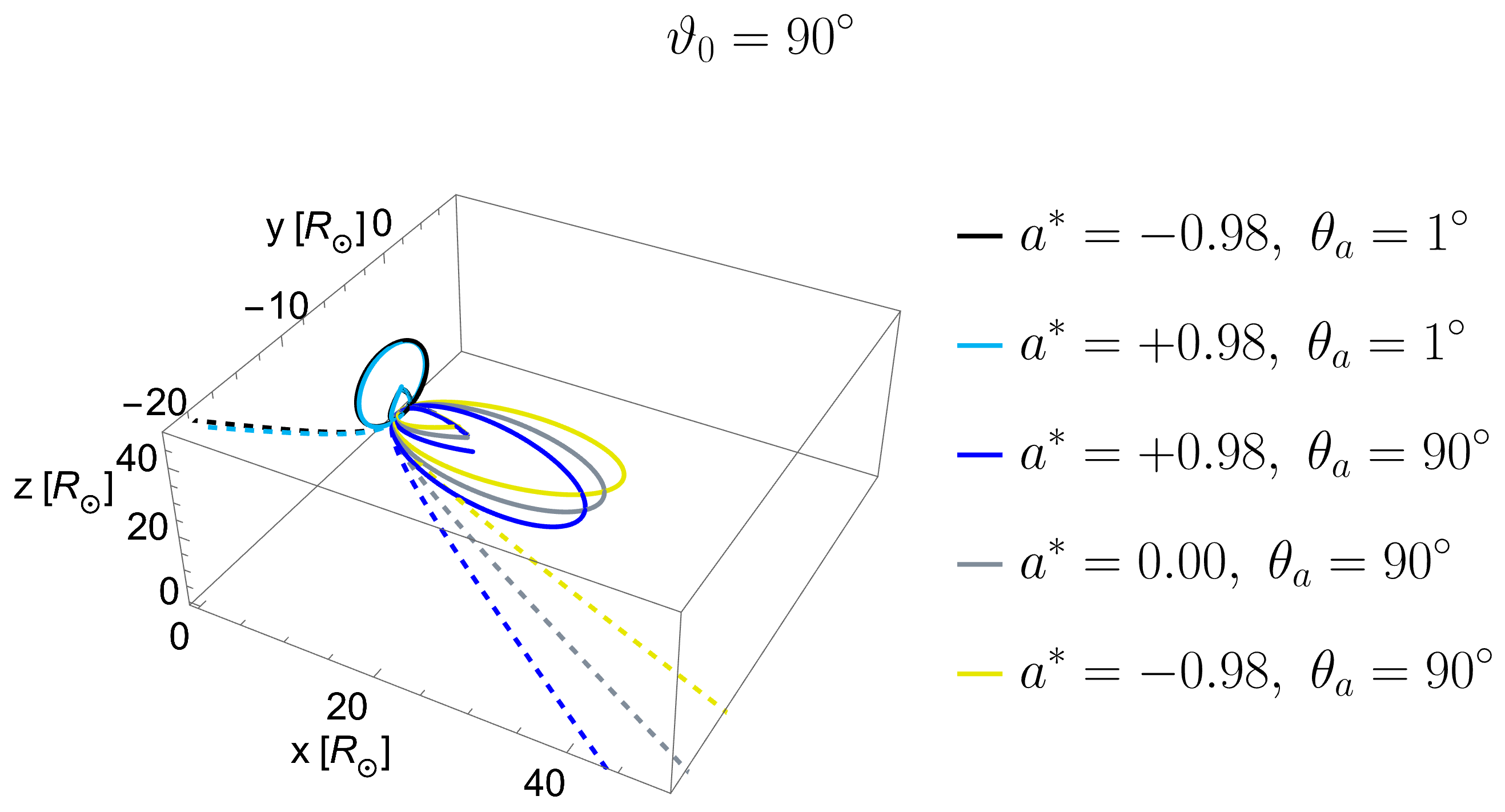}
	\caption{{\small Trajectories of binary test particle around the spinning BH whose initial relative angular momenta of binary is subtended by $\vartheta_0=90^\circ$. \tb{The solid and dashed lines denote the point-masses $m_1$ and $m_2$ respectively. The trajectories are plotted for both spinning and non-spinning (Schwarzschild) BH. For spinning BH, equatorial and off-equatorial orbits are displayed corresponding to $\theta_a=90^\circ$ and $\theta_a=1^\circ$ respectively with BH spin parameter $\astar=\pm0.98$. The different colors correspond to the different BH spin parameters and initial polar angles indicated in the legend.}}}
	\label{xyz2}
\end{figure}

If the initial binary inclination differs from the complete prograde or retrograde rotations, the motion is no longer expected to remain restricted to the orbital plane. As $\vartheta_0$ increases from $0^\circ$ toward $90^\circ$, we find that the tidal separation between the particle trajectories reduces following the Hills mechanism at pericentre, and the resulting bound elliptical orbits become progressively more compact. Figure \ref{xyz2} shows the trajectories for the case $\vartheta_0 = 90^\circ$, where the inner orbital motion of the binary is no longer aligned with its outer orbital motion. The resulting dynamics of the tidal encounter are qualitatively similar to those of the prograde rotation case. However, in $\vartheta_0=90^\circ$ case, the particle is caught into a smaller and more tightly bound elliptical orbit. For example, the outer orbit of particle 1 has a semi-major axis $a_1^{\mathrm{out}} \simeq 2180~r_g$ (where $r_g=0.017~R_{\odot}$) in the prograde rotation case, whereas it reduces to $a_1^{\mathrm{out}} \simeq 1590~r_g$ for $\vartheta_0=90^\circ$. For higher initial binary inclinations ($\vartheta_0 > 90^\circ$), the Hills mechanism is progressively suppressed, and the tidal separation continues to decrease, allowing the binary to remain bound through pericentre passage. In these cases, the particles depart from the inner orbital motion at much later times, leading to highly eccentric bound orbits around the BH. Eventually, at $\vartheta_0 = 180^\circ$, the binary components never detach from their inner orbital motion, and the Hills mechanism ceases to operate entirely.

\begin{figure}[h]
	\epsscale{1.15}
	\plottwo{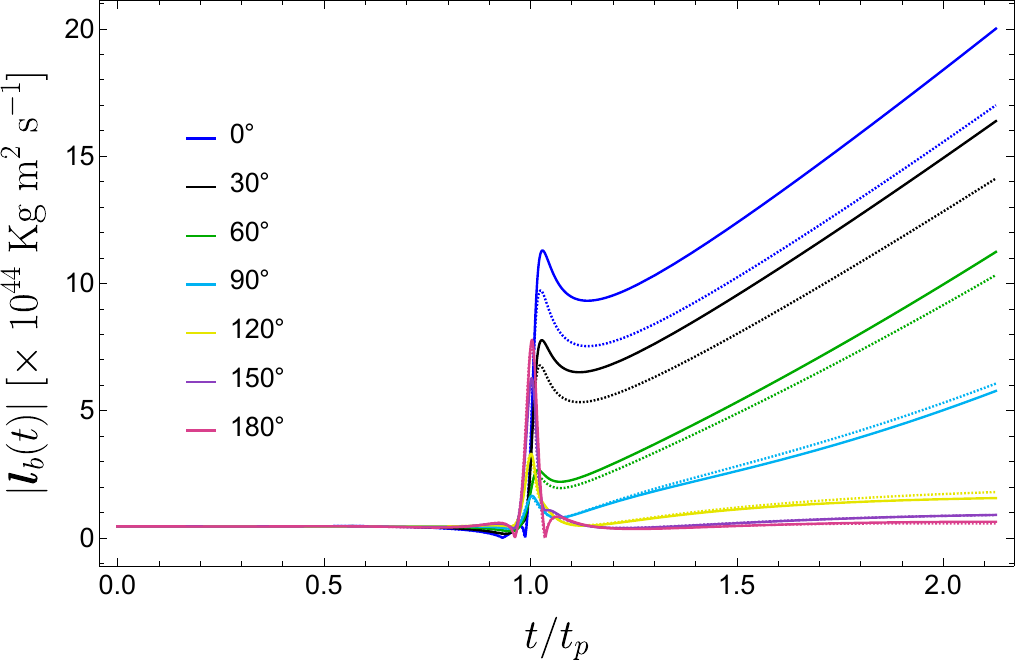}{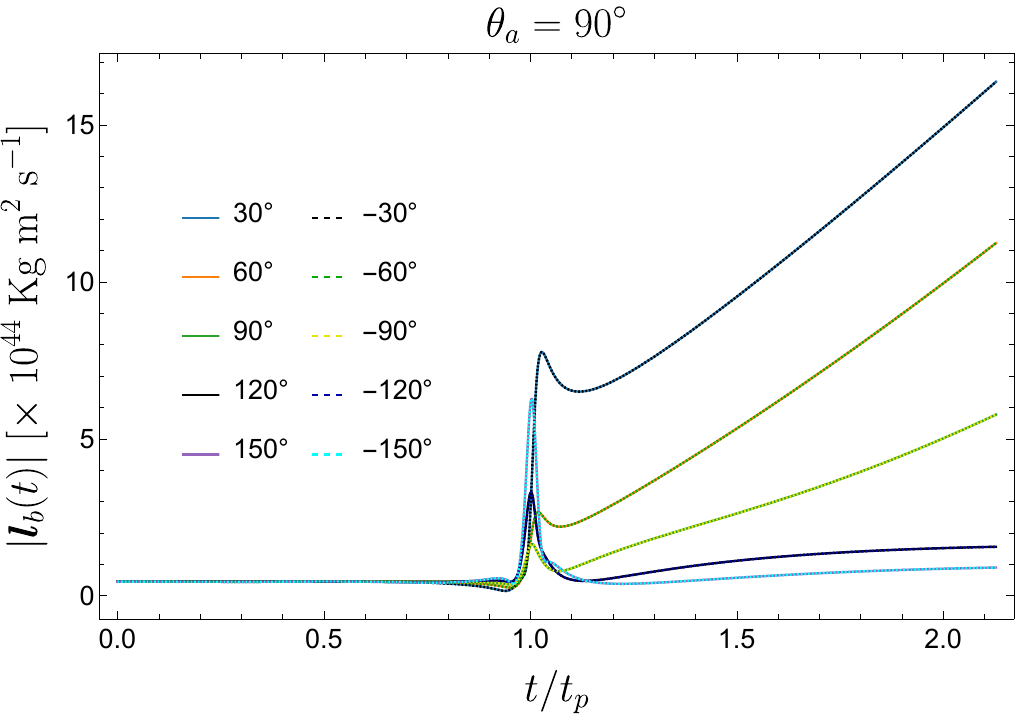}
	\caption{{\small Evolution of the binary spin around a spinning BH ($\astar=-0.98$). \textbf{Left Panel: }The magnitude of the binary spin, $|\boldsymbol{l}_b(t)|$ over normalized time ($t/t_p$, where $t_p=0.0624~\mathrm{hr}$ is the pericentre time) for different initial binary inclinations $\vartheta_0$ around spinning BH. There is abrupt gain in the magnitudes of binary spin at $t_p$ from its initial value, $|\boldsymbol{l}_b(0)|=|\mathbf{L}_b|$. The solid lines denote the values of $|\boldsymbol{l}_b(t)|$ in equatorial orbit, $\theta_a=90^\circ$, and dashed lines denotes off-equatorial orbit, $\theta_a=1^\circ$. \textbf{Right panel:} \tb{The magnitude of the binary spin, $|\boldsymbol{l}_b(t)|$ over normalized time ($t/t_p$) for equatorial orbits, $\theta_a=90^\circ$ corresponding to initial binary inclinations $\vartheta_0$ (solid lines) and $-\vartheta_0$ (dashed lines).  The curves coincide for each pair of opposite inclinations, indicating the corresponding degeneracy.} }}
	\label{binary_spin}
\end{figure}

To understand these effects, we can examine how the binary spin is affected by the tidal torque exerted during the encounter with BH for different 
inclinations $\vartheta_0$. To this end, we compute the magnitude of the binary’s inner angular momentum (spin), \tb{$|\boldsymbol{l}_b(t)|$} for a range of initial inclinations.
\tb{Here, $\boldsymbol{l}_b$ denotes the same binary inner angular momentum $\mathbf{L}_b$ introduced in Section 2, such that $|\boldsymbol{l}_b(0)|=|\mathbf{L}_b|$.} 
We vary $\vartheta_0$ from $0^\circ$ to $180^\circ$ for BH spin $\astar=-0.98$ and $\theta_a=1^\circ$, $90^\circ$ (a similar analysis can also be carried out 
for $\astar=0.98$, but we omit this for brevity). In the left panel of Figure \ref{binary_spin}, we find that the magnitude of the binary spin diverges after pericentre passage for inclinations  in the range $0^\circ \le \vartheta_0 \lesssim 90^\circ$. For larger inclinations, $90^\circ < \vartheta_0 \le 180^\circ$, the spin magnitude remains finite. 
This implies that the Hills mechanism operates most strongly for binaries whose initial inclinations lie between $0^\circ$ and $90^\circ$. 
For larger inclination angles (particularly approaching towards retrograde rotation), the binary tends to retain its bound state more efficiently, and the tidal separation continues to decrease as the system transitions from prograde to retrograde configurations. We further find that the gain in the peak value of the binary spin magnitude is maximal for prograde configurations, decreases to a minimum near $\vartheta_0 = 90^\circ$, and then increases again toward the retrograde limit. This trend mirrors the behaviour of the binding energy of the particle captured during the Hills mechanism, suggesting a close correlation between the spin gain and the bound nature of the post-encounter orbit. Another important point is that the magnitude of the binary spin exhibits a significant difference between $\theta_a=1^\circ$ and $90^\circ$ cases for prograde rotation ($\vartheta_0 \simeq 0^\circ$). This difference gradually drops as $\vartheta_0$ increases, and the spin magnitudes converge in the retrograde limit. Thus, the tidal torque acting on the binary leads to a significantly large gain in the spin magnitude for equatorial encounters ($\theta_a = 90^\circ$) than for off-equatorial ones ($\theta_a = 1^\circ$) when the configuration is prograde rotation. In contrast, for retrograde configurations ($\vartheta_0 \simeq 180^\circ$), the gain is nearly same. 

\begin{figure}[h]
	\epsscale{1.15}
	\plottwo{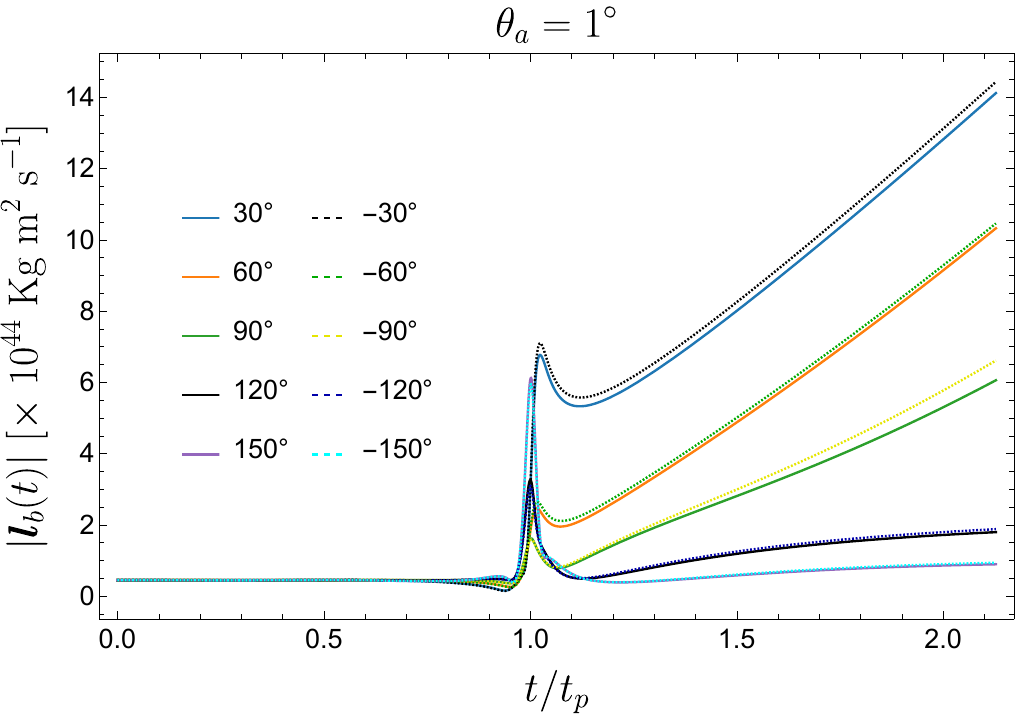}{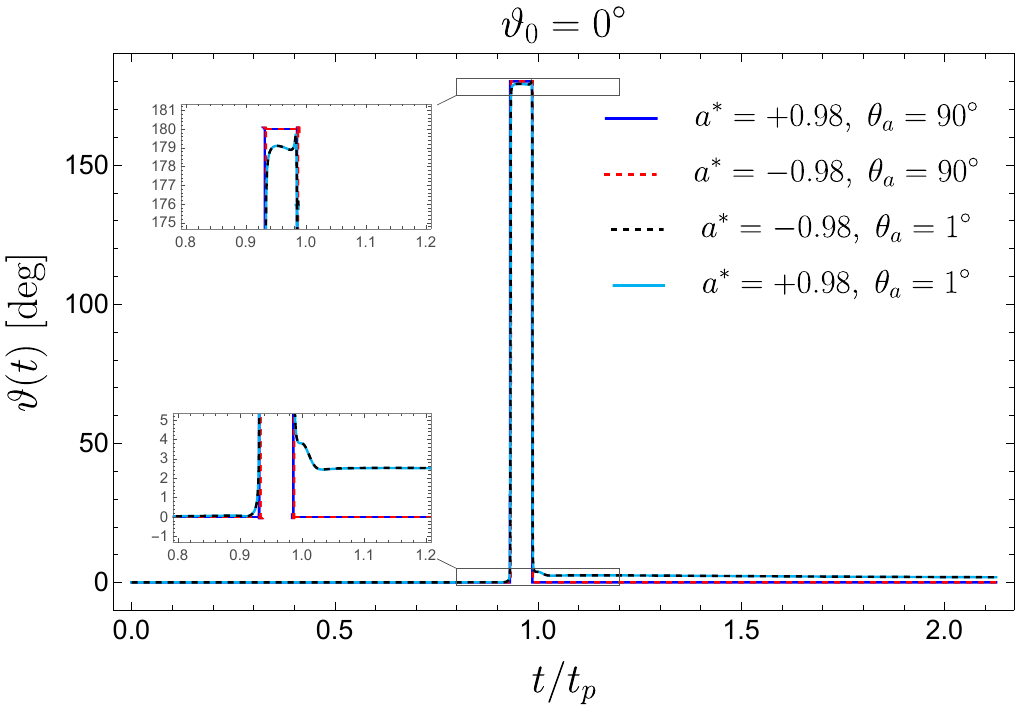}
	\caption{{\small \tb{Evolution of the binary spin around a spinning BH. \textbf{Left Panel: }The magnitude of the binary spin, $|\boldsymbol{l}_b(t)|$ over normalized time ($t/t_p$) for off-equatorial orbits, $\theta_a=1^\circ$ and BH spin $a^\star=-0.98$, corresponding to initial binary inclinations $\vartheta_0$ (solid lines) and $-\vartheta_0$ (dashed lines).  The curves do not coincide for each pair of opposite inclinations, indicating the breaking of the degeneracy. \textbf{Right Panel: } Reorientation of binary spin over time. $\vartheta(t)$ measures the orientation of $\boldsymbol{l}_b$ with respect to $\mathbf{L}_o$. The direction of the initial binary spin, $\mathbf L_b$, corresponds to $\vartheta_0=0^\circ$ inclination.The dashed, and solid curves correspond to BH spin $a^\star=+0.98$ and $a^\star=-0.98$, respectively.} }}
	\label{binary_spin_2}
\end{figure}

Now, to explore the remaining initial binary inclinations, we can consider  configurations related by $\vartheta_0 \to -\vartheta_0$. For a non-spinning BH, these cases are redundant owing to the inherent symmetry in the relative alignment between the binary spin and its OAM. In this situation, we obtain identical outcomes for every pair $\vartheta_0$ and $-\vartheta_0$, since both correspond to the same initial relative orientation of the binary spin with respect to the OAM. This symmetry also persists for equatorial orbits around a spinning BH, leading to identical outcomes (\tb{see the right panel of Figure \ref{binary_spin}}), because the BH spin always remains aligned with the OAM. However, for off-equatorial orbits, the BH spin breaks this symmetry. \tb{As shown in the left panel of Figure \ref{binary_spin_2}}, small but measurable differences arise for $\vartheta_0$ and $-\vartheta_0$ due to the frame-dragging effects associated with the spinning BH in the $\theta_a = 1^\circ$ configuration. Consequently, our interest lies mainly within the range $0^\circ \le \vartheta_0 \le 180^\circ$ in this study.

\begin{figure}[h]
	\epsscale{1.15}
	\plottwo{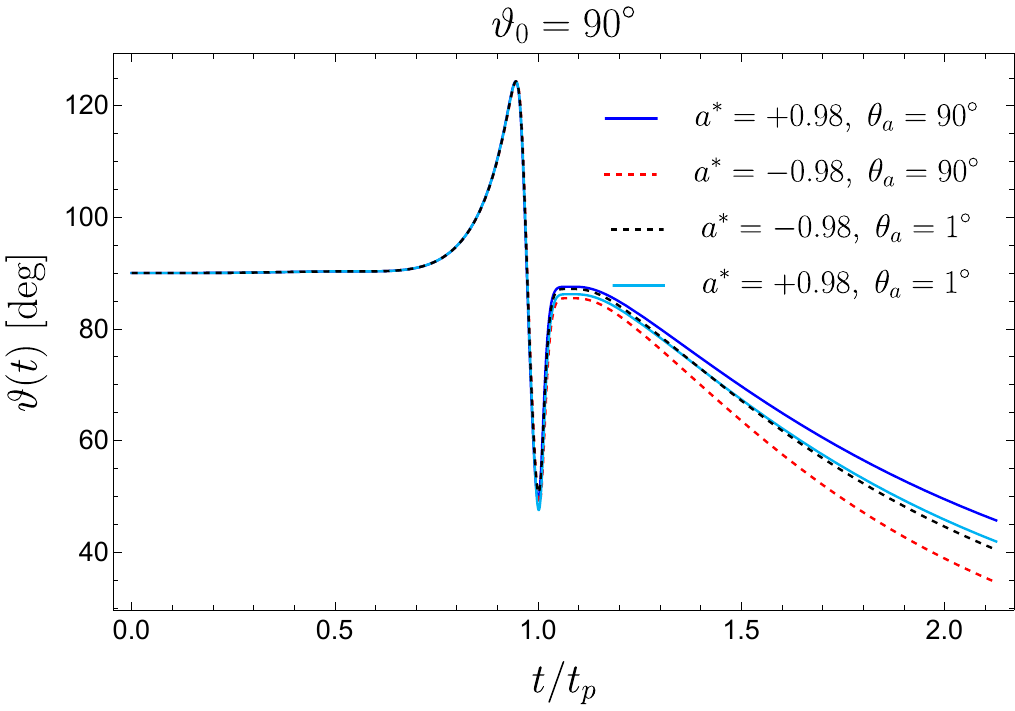}{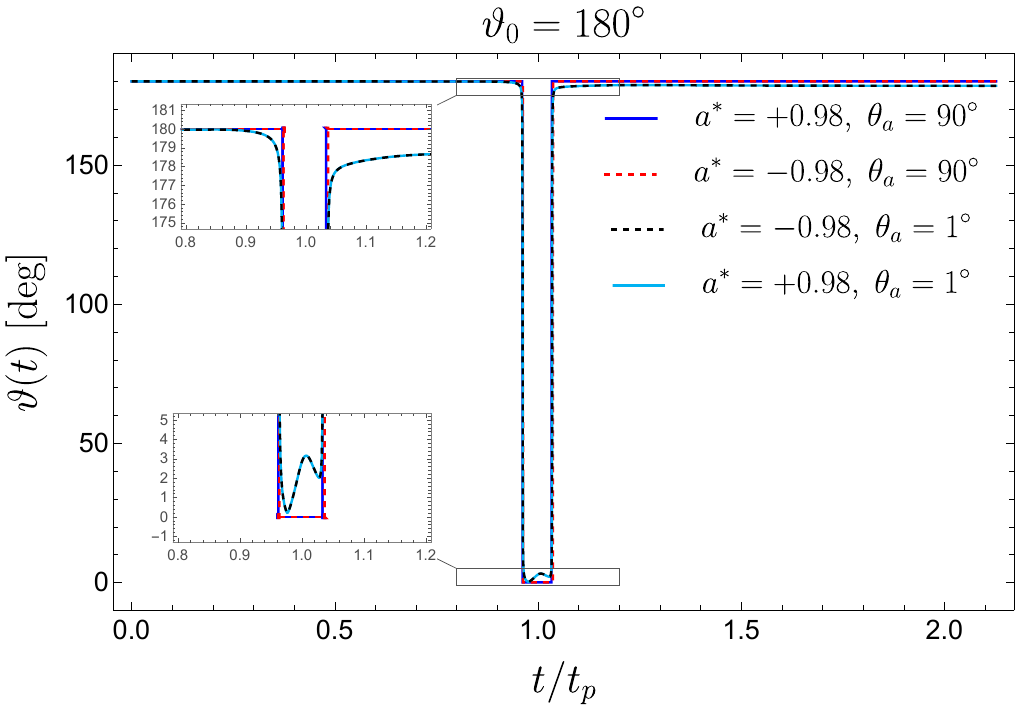}
	\caption{{\small \tb{Reorientation of binary spin over time. $\vartheta(t)$ measures the orientation of $\boldsymbol{l}_b$ with respect to $\mathbf{L}_o$. The dashed, and solid curves correspond to BH spin $a^\star=+0.98$ and $a^\star=-0.98$, respectively. The direction of the initial binary spin, $\mathbf L_b$, corresponds to \textbf{Left Panel: }$\vartheta_0=90^\circ$ inclination and \textbf{Right Panel: }$\vartheta_0=180^\circ$ inclination.} }}
	\label{binary_spin_3}
\end{figure}

In these encounters, the tidal torque also alters the initial alignment of the binary’s spin. We assess how the spin undergoes an abrupt reorientation near pericentre, including the influence of the BH spin. \tb{In the right panel of Figure \ref{binary_spin_2}, the left panel of Figure \ref{binary_spin_3} and the right panel of Figure \ref{binary_spin_3}, we show the evolution of the relative orientation of the binary spin, with respect to its initial OAM $\mathbf{L}_o$ for three inclinations, $\vartheta_0 = 0^\circ$, $90^\circ$, and $180^\circ$, respectively, and for different $\astar$ and $\theta_a$.}  We find that the binary experiences an almost complete orbital flip in both the prograde ($\vartheta_0=0^\circ$) and retrograde ($\vartheta_0=180^\circ$) rotations. Similar flipping effect of the binary spin has been analytically examined in earlier works by \cite{Will2017}, \cite{2025PhRvD.111h3052C}, using higher-order multipole expansions of the Newtonian potential incorporating PN and relativistic corrections. However, in our case of spinning BH, where the dynamics follow the full Kerr geometry, we find that the orbital flip is less stable for the off-equatorial orbit than the equatorial one. \tb{For instance, as shown in the right panel of Figure \ref{binary_spin_3}, corresponding to $\theta_a=1^\circ$ orbit, the dashed black and solid cyan curves (identical for both $\astar = \pm 0.98$) barely reach $0^\circ$, and also they do not retain their initial alignment fully after pericentre passage. They exhibit a maximum relative deviation of $\sim 3^\circ$ compared to the equatorial case. Same can be seen for the prograde rotation (see right panel of Figure \ref{binary_spin_2}). For the $\vartheta_0 = 90^\circ$ case (left panel Figure \ref{binary_spin_2}), the largest deviations from the initial spin orientation after $t=t_p$ occur for the $\astar = -0.98$, $\theta_a = 90^\circ$ configuration, compared to the other cases, although all cases eventually tend to align with the BH orbital plane ($\vartheta(t)\to 0^\circ$).} These relative deviations can be attributed to the overall Lense–Thirring precession induced by the spinning BH in our setup. As is well known, around a spinning BH the Lense–Thirring effect induces nodal precession for off-equatorial orbits \citep{1918PhyZ...19..156L, 1972ApJ...178..347B}. Consequently, when a rotating object orbits a spinning BH, it undergoes a wobbling motion. In our case, although the encounter occurs on a parabolic orbit, this single interaction is enough to introduce an overall instability in the orbital flip of binary spin corresponding to the off-equatorial orbit.

\subsection{Dynamics due to the coupling effect of angular momenta}\label{spin_coupling}

\begin{figure}[h]
	\epsscale{1.11}
	\plottwo{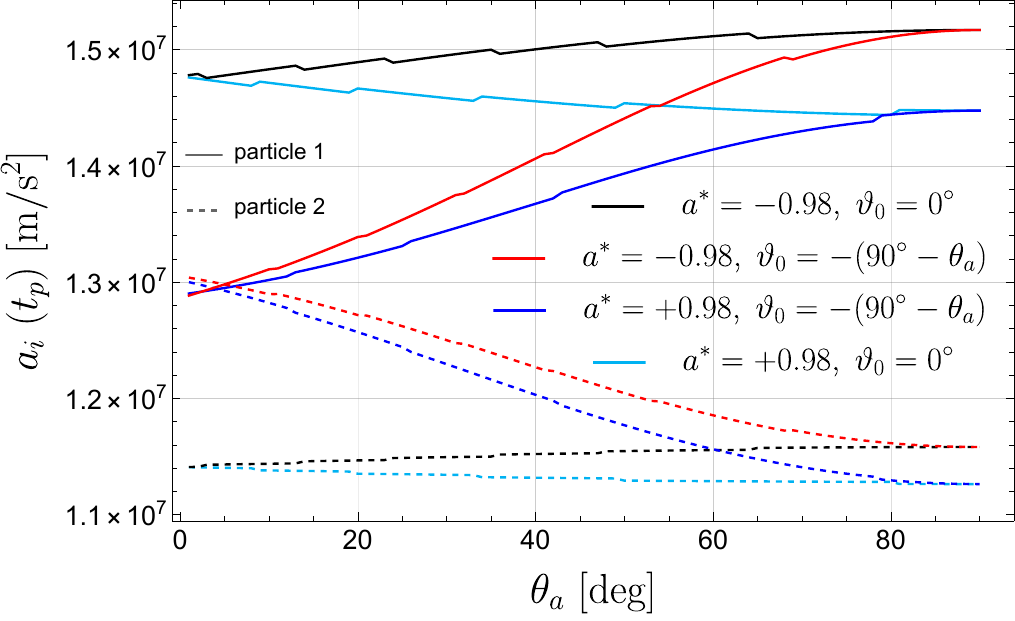}{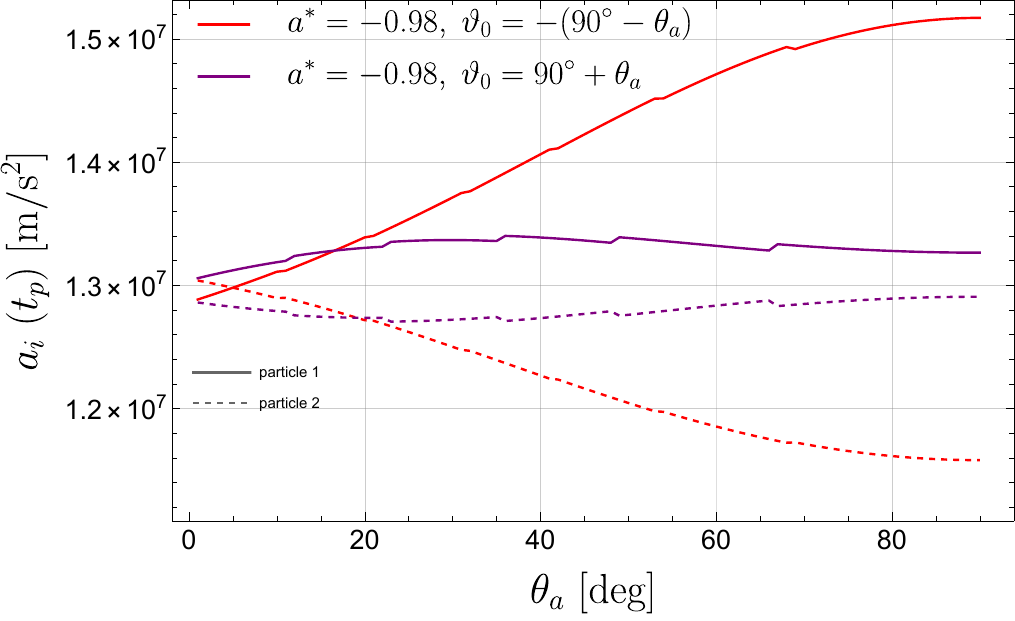}
	\caption{{\small Comparison of acceleration magnitudes exerted on each individual particle between when binary spin - BH spin is coupled and binary spin - OAM is coupled. }}
	\label{acc}
\end{figure}

To investigate the coupling of binary spin with BH spin and/or with its OAM around the BH, we briefly account the resultant dynamics near pericentre. We compute the magnitude of the accelerations of each particle at pericentre time $t_p$ for different values of $\theta_a$, as shown in Figure \ref{acc}. In the left panel, we compare the couplings: one in which the binary spin is aligned with its OAM ($\vartheta_0 = 0^\circ$), and another in which the binary spin is aligned with the $z$-axis ($\vartheta_0 = \theta_a-90^\circ$). We find that the differences arise across different off-equatorial orbits, however, the differences for each individual particles do not grow by the same amount. Moreover, the accelerations in prograde rotations, increases for $\astar=-0.98$ as we go towards the equatorial orbit, while it decreases for $\astar=+0.98$ spin. This is unlike the behavior in $\vartheta_0=-(90-\theta_a)$ motion, where the acceleration either increases or decreases for both $\astar=\pm 0.98$.  To highlight the relative strength of these couplings, we focus on the cases $\theta_a = 1^\circ$ and $\theta_a = 90^\circ$, where the differences are substantial for both $\astar = \pm 0.98$. Further, the outcomes in the coupling between the BH spin and the binary spin depends on whether the binary spin is aligned ($\vartheta_0 = \theta_a-90^\circ $) or anti-aligned ($\vartheta_0 = 90^\circ + \theta_a$) with the $z$-axis. We therefore consider both possibilities and the resulting accelerations are compared (for $\astar = -0.98$) in the right panel. All these comparisons are meaningful only if we fix the binary phase to $\varphi_0 = 0^\circ$. Indeed, from a similar computation we found that most other values of $\varphi_0$ do not retain this relative differences in the strength of couplings. Therefore, the combination  $\varphi_0 = 0^\circ$ with $\theta_a = 1^\circ$ and $\theta_a = 90^\circ$ constitutes a reasonable set of initial conditions that yields maximally distinguishable outcomes in our analysis. Accordingly, this combination is adopted as the representative setup for the analyses that follow.



\section{Dynamics of binary including hydrodynamics and finite stellar sizes}

Now we present the results of our numerical simulations. As mentioned before, we integrate the Kerr geodesic equations, while the self-gravity and hydrodynamics of the WDs are treated in the Newtonian approximation. \tb{The validity of this treatment can be assessed by comparing the characteristic sizes of the WDs and the binary separation with the local curvature scale of the external Kerr spacetime, as discussed at the beginning of Section \ref{sec:binary_spin}. Using Equation \ref{alpha}, we find that the external curvature varies only weakly across the individual WDs and the binary at the encounter location. In addition, the individual WDs are themselves only weakly self-gravitating, with compactness $GM_{\rm WD}/(R_{\rm WD}c^2)\simeq7.5\times10^{-5}$, so that relativistic corrections to the internal stellar structure and to the WD's own gravitational field are negligible. The importance of relativistic corrections to the mutual gravitational interaction is further quantified by the 1PN comparison presented in Appendix \ref{appendix1}. Finally, in Appendix \ref{appendix:SPH_velocity}, we examine the Lorentz factor of all SPH particles at key dynamical stages to verify that the stellar material remains non-relativistic. Thus, while the mutual gravitational interaction and hydrodynamic evolution of the stellar material are treated in the Newtonian regime, each SPH particle is individually evolved in the global frame under the external Kerr spacetime, thereby accounting for the relativistic orbital motion induced by the central BH.} We have run a suite of 18 SPH simulations in our study. The number of 
SPH particles is $3\times 10^5$ particles per each WD. \tb{We have confirmed
	numerical convergence by repeating most of these simulations with
	$10^5$ and $5\times10^5$ particles per WD. A few convergence
	tests are presented in Appendix~\ref{appendix:res}}.

\subsection{Mass disruption}

\begin{figure}[h]
	\epsscale{1.17}
	\plottwo{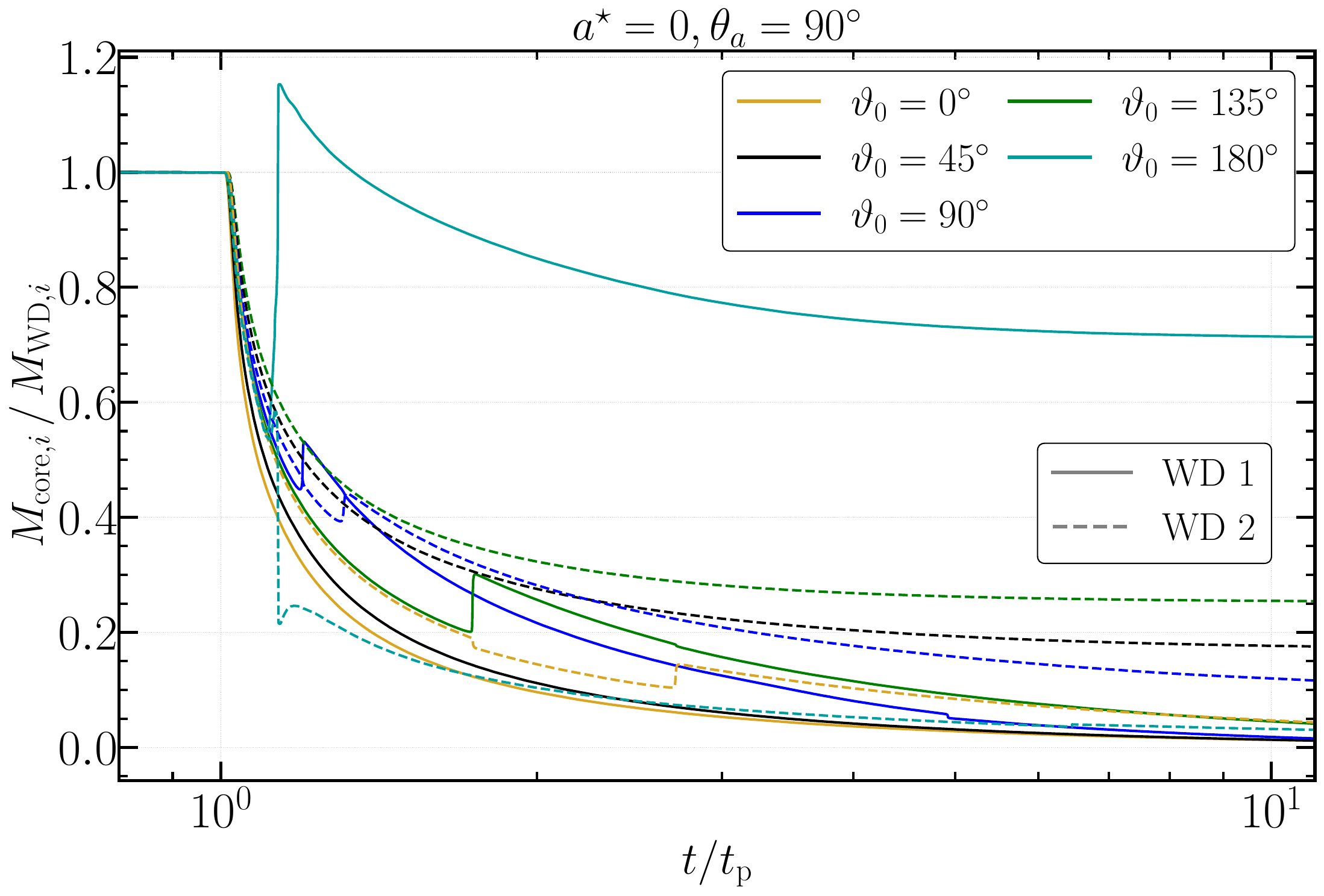}{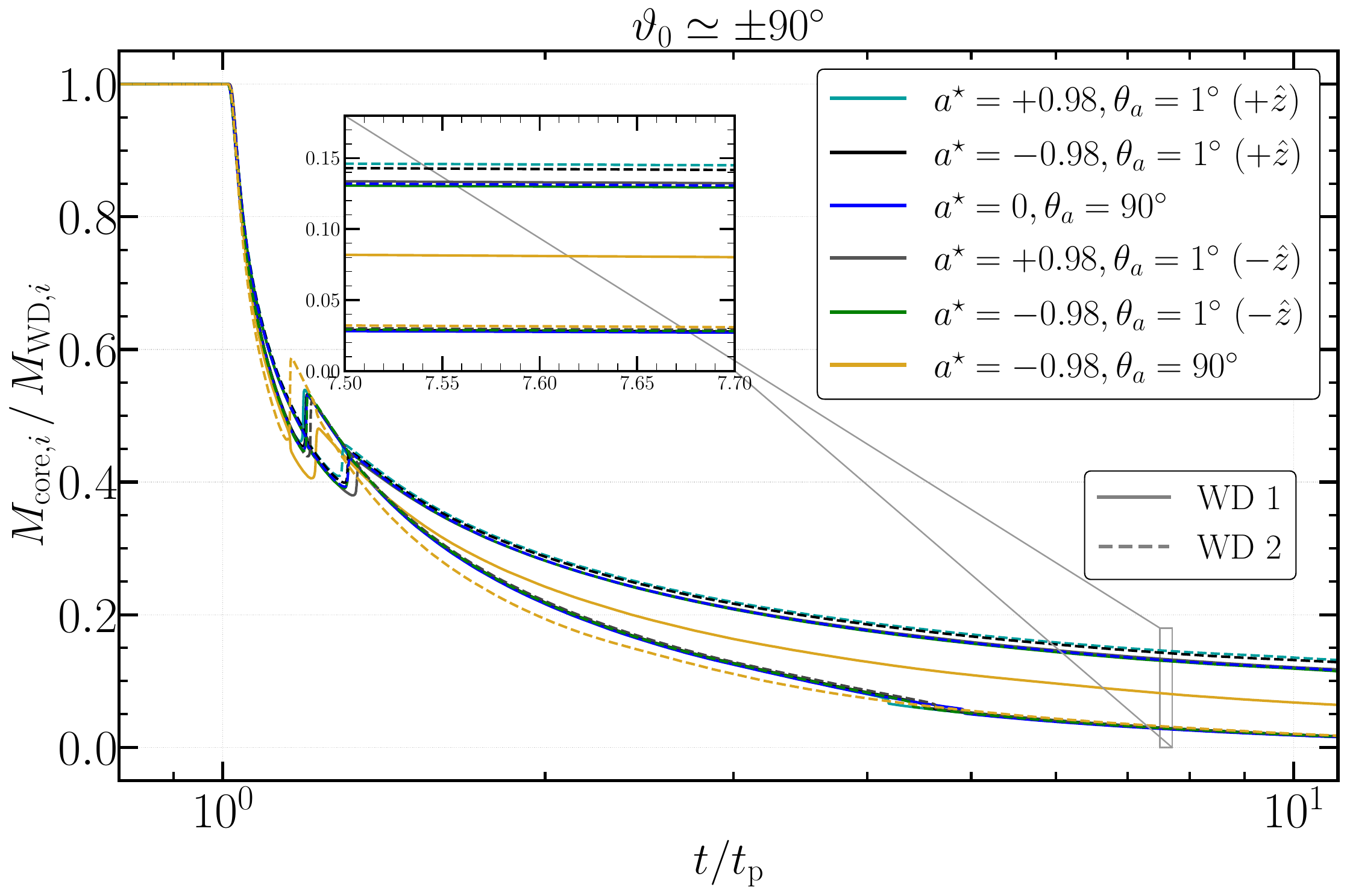}
	\caption{{\small Bound core mass fraction, \( M_{{\rm core},i} / M_{{\rm WD},i} \), for each WD as a function of \( t / t_p \), where \( i = 1, 2 \) and \( t_p \approx 0.062\,\mathrm{hr} \) marks the time of pericenter passage. 
			\textbf{Left panel:} Evolution for a non-spinning (Schwarzschild) BH for various initial binary inclinations \( \vartheta_0 \) and phase $\varphi_0=0^{\circ}$. 
			\textbf{Right panel:} Evolution in the presence of BH spin, with the initial binary spin either aligned or anti-aligned with the \( z \)-axis ($\vartheta_0 \approx \pm 90^{\circ}$) for the \( \theta_a = 1^{\circ} \) orbit. 
			For comparison, the cases \( a^{\star} = 0,\, \vartheta_0 = 90^{\circ} \) and \( a^{\star} = -0.98,\, \theta_a = -\vartheta_0 = 90^{\circ} \) are included, highlighting their similarity to the \( \theta_a = 1^{\circ} \) and $\vartheta_0 \approx \pm 90^{\circ}$. An inset highlights regions where multiple cases overlap, forming three distinct bands: (i) cyan and black (dashed), (ii) grey and green (solid) with blue (dashed), and (iii) the complementary set consisting of cyan and black (solid), grey and green (dashed), and blue (solid).
	}}
	\label{mbound_az}
\end{figure}

The core mass fraction is calculated following the method described in \cite{binary1}. We first examine the case \(a^\star = 0\) and vary the binary inclination \(\vartheta_0\); the corresponding results are shown in left panel of  Fig.~\ref{mbound_az}. Throughout this analysis, we adopt an equatorial orbit\footnote{In Schwarzschild spacetime, however, any orbital plane can be transformed into an equivalent equatorial configuration; therefore, the results are independent of the orbital plane.}. As the binary inclination is varied, the degree of tidal disruption experienced by WD1 and WD2 changes markedly. For inclinations up to \(\vartheta_0 = 90^{\circ}\), one star is completely disrupted while the other undergoes only partial disruption. At \(\vartheta_0 = 135^{\circ}\), both stars are fully disrupted. Once the orbital motion becomes retrograde, the roles of the two stars reverse: the star previously experiencing partial disruption is instead fully disrupted, consistent with expectations from the initial orbital geometry.

A notable feature in the retrograde case is the pronounced interaction between the surviving stellar cores, which leads to mass exchange between them. Core--core mass transfer is also evident at \(\vartheta_0 = 135^{\circ}\), where an increase in one core's mass corresponds to a decrease in the other. In contrast, at \(\vartheta_0 = 90^{\circ}\), the mass transfer proceeds from the tidal tail of one WD into the core of the other, as indicated by the rise of one core-mass curve without a corresponding decrease in the other. We also see here (as demonstrated in section \ref{sec:binary_spin}) that the Hills mechanism indeed produces exchange-type interactions particularly for inclination angles up to $\vartheta_0=90^\circ$. Also, in $\vartheta_0=90^\circ$ case, we see inherent interactions develops among the disrupted materials of both WDs, despite the Hills mechanism driving the disrupted core to eject in a parabolic trajectory together with the disrupted tails. \tb{We illustrate this further with close-range snapshots of the simulation when discussing the fallback rates below.} This is because, the tidal separation between WDs in $\vartheta_0=90^\circ$ is quite less than the $\vartheta_0=0^\circ$ case (see left panel Figure \ref{xyz1} and Figure \ref{xyz2}). For larger inclinations, the inherent interactions become more dominant due to the bound nature of binary, and the behaviour is primarily governed by the dynamics of the disrupted material of bound pair. 

\begin{figure}[h]
	\epsscale{1.17}
	\plottwo{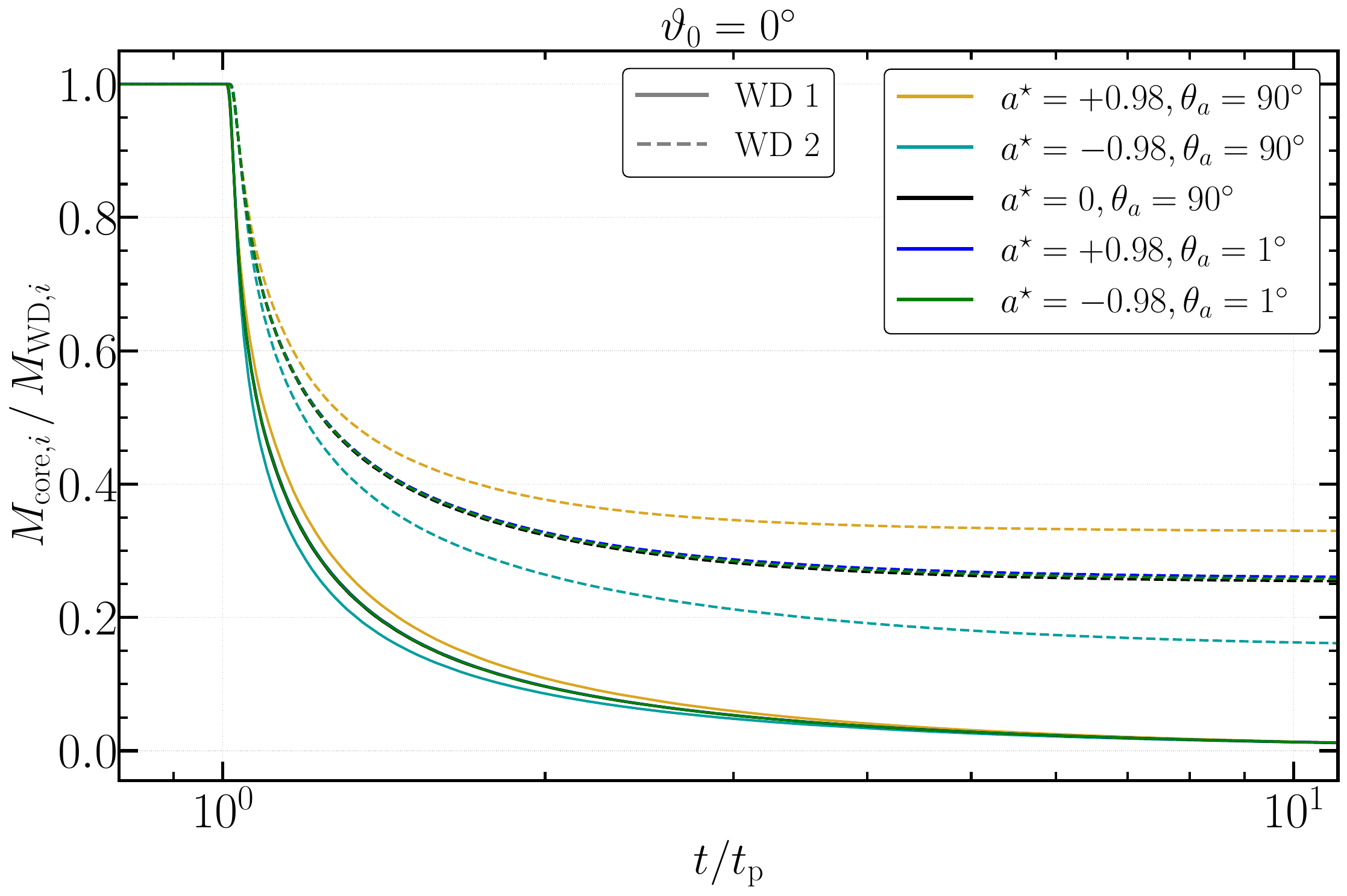}{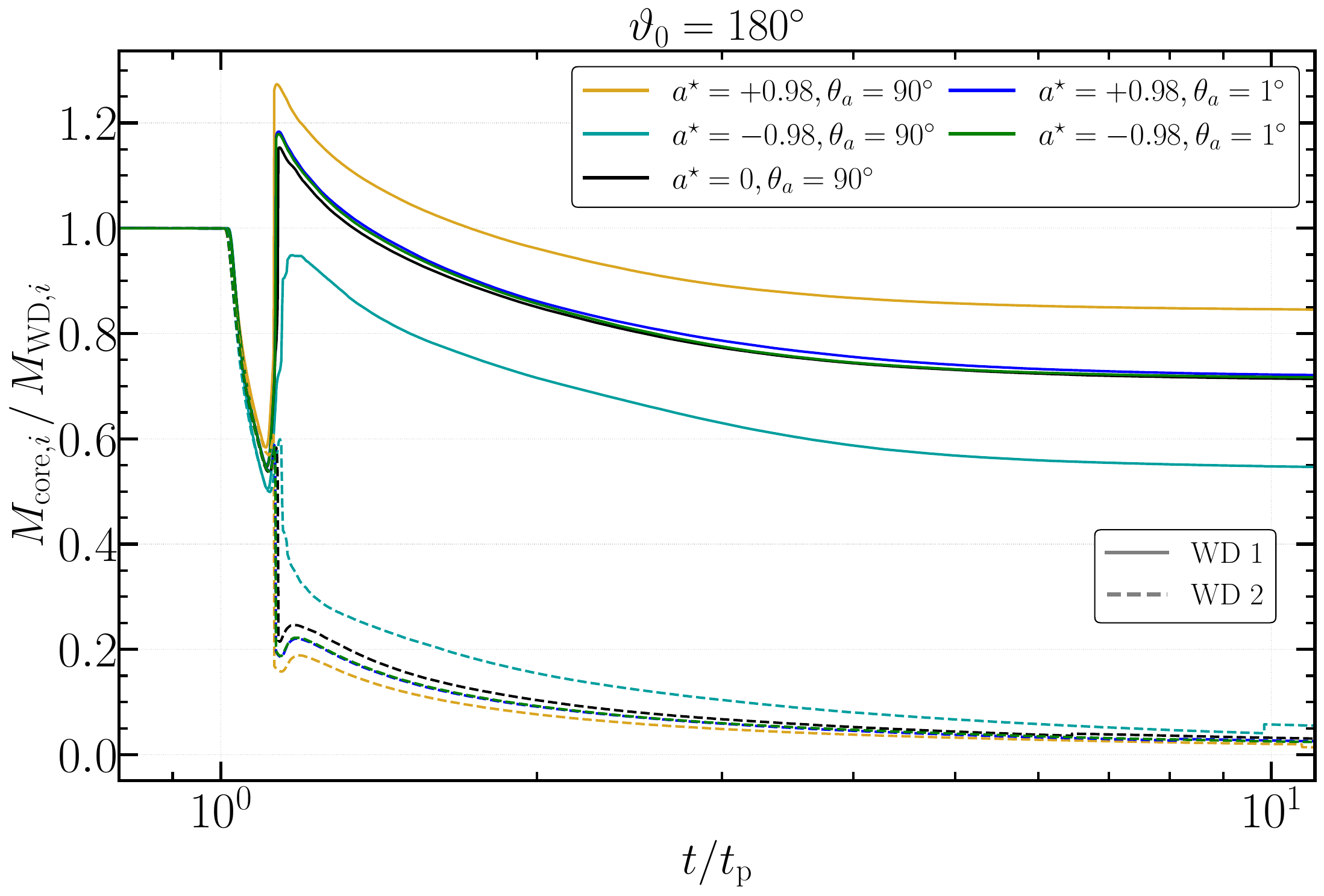}
	\caption{{\small Bound core mass fraction, \( M_{{\rm core},i} / M_{{\rm WD},i} \), for each white dwarf as a function of \( t / t_p \), where \( i = 1, 2 \) and \( t_p \approx 0.062\,\mathrm{hr} \) denotes the time of pericenter passage. 
			\textbf{Left panel:} Prograde binaries ($\vartheta_0=0^\circ$) orbiting a spinning BH, shown for both equatorial and off-equatorial orbits. 
			\textbf{Right panel:} Retrograde binaries ($\vartheta_0=180^\circ$) orbiting a spinning BH, likewise shown for equatorial and off-equatorial orbits. 
			For comparison, the corresponding $a^{\star}=0$ (non-spinning BH) prograde and retrograde cases are included in each panel.}}
	\label{mbound_pr}
\end{figure}

We next investigate the coupling between the binary spin and the BH spin, as well as the coupling between the binary spin and the OAM. For this purpose, we consider rapidly spinning BH with \(a^\star = +0.98\) and \(a^\star = -0.98\), for both equatorial and off-equatorial orbit, and for binary spins that are initially prograde or retrograde. The left panel of Figure~\ref{mbound_pr} shows the prograde binary spin configurations. We find that the introduction of a positive BH spin (\(a^\star = +0.98\)) reduces the disruption experienced by WD2 compared to the Schwarzschild case, while a negative BH spin (\(a^\star = -0.98\)) enhances the disruption. For off-equatorial orbits with a small tilt (\(\theta_a = 1^{\circ}\)), the degree of disruption is similar to the \(a^\star = 0\) case for both signs of the BH spin, lying between the two equatorial-spin extremes.

These results qualitatively resembles the behavior observed in the case of a single spinning WD orbiting an spinning IMBH, as discussed extensively in \cite{Garain3}. In the retrograde binary-spin configurations (right panel of Figure~\ref{mbound_pr}), interaction between the stellar cores again leads to alternating increases and decreases in core mass. This time rather than disruption, we obtain the differences in the saturated mass of the WD1 by gaining material from its companion due to inherent interactions of the retrograde binary configuration. Nevertheless, the final saturated core mass of WD1 across different configurations follows the same qualitative trend observed for WD2 in the prograde case.

An especially interesting result arises in the right panel of Figure~\ref{mbound_az}, which shows the core-mass evolution for orbits with \(\theta_a = 1^{\circ}\) and binary spin lie along the \(\pm z\) direction (which corresponds to almost a perpendicular type configuration in $\vartheta_0$). In these configurations, the core masses closely track the \(a^\star = 0\), \(\vartheta_0 = 90^{\circ}\) (binary spin lies in the orbital plane and points to the BH) Schwarzschild case. This behavior indicates that when the orbit is parallel to the BH spin (i.e., the OAM is perpendicular to the BH spin), the influence of BH spin is effectively suppressed, and the system behaves as if the BH were non-spinning. Among the spinning and non-spinning BH cases, the maximum disruption happens for \(a^\star = -0.98\), \(\theta_a=-\vartheta_0 = 90^{\circ}\) (binary spin lies in the orbital plane and points away from the BH) case.

\subsection{Tidal Stretching}
\begin{figure}[h]
	\epsscale{1.1}
	\plottwo{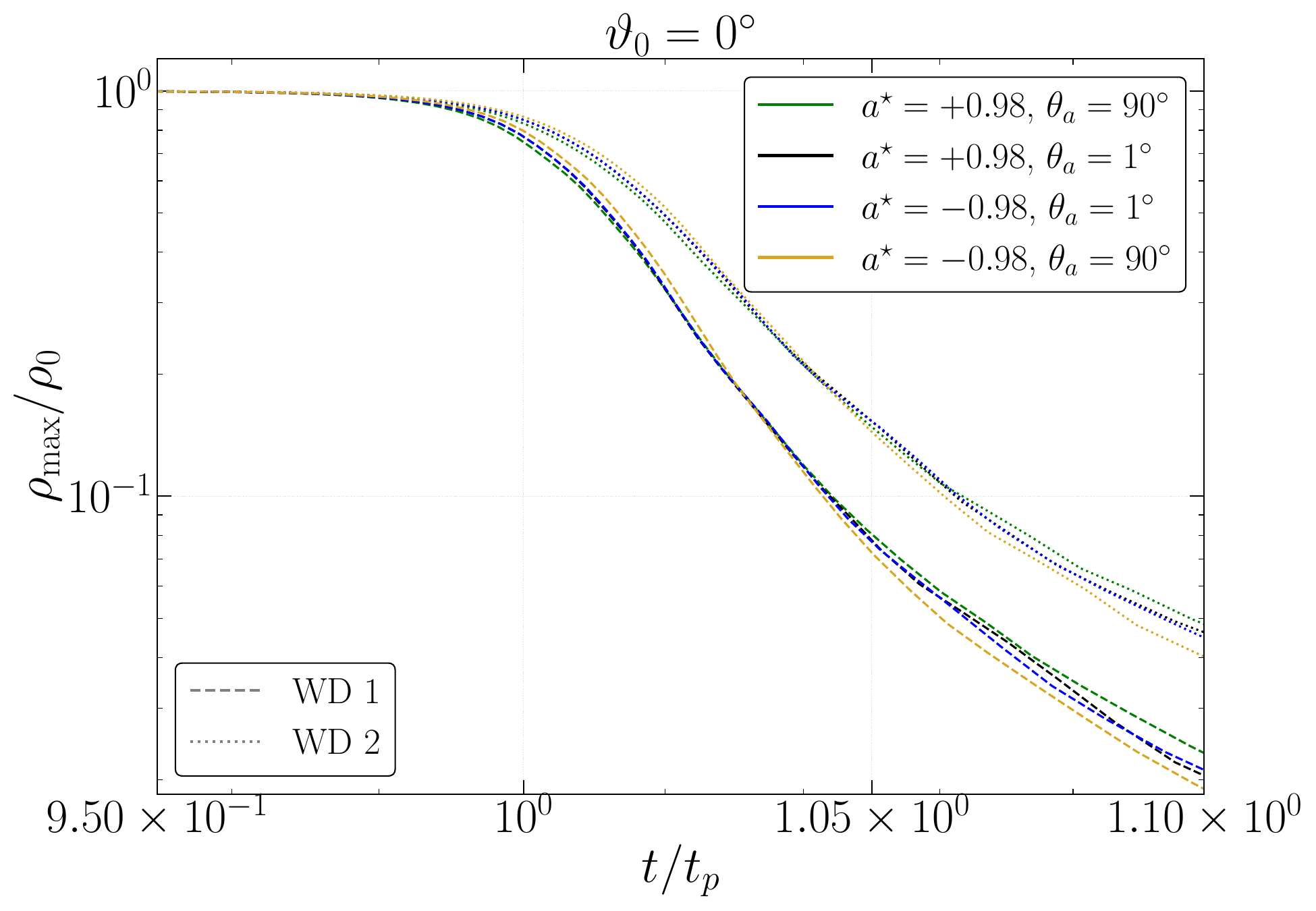}{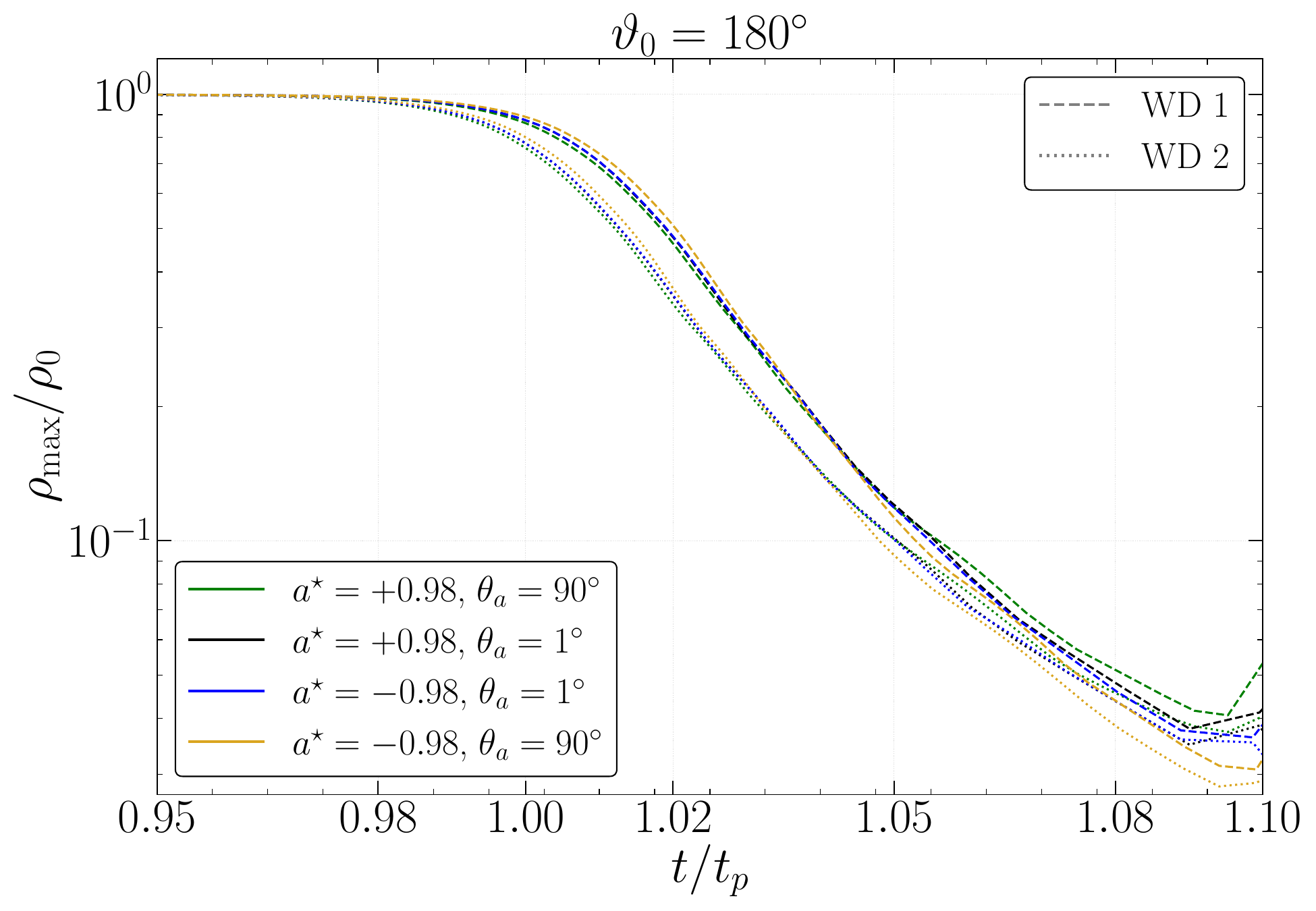}
	\caption{{\small Evolution of the maximum (central) density of each WD, $\rho_{\max}$,
		scaled by its initial value $\rho_0$, as a function of normalized time $t/t_p$ around
		pericentre passage ($t_p \approx 0.062$\,hr).
		Dashed and dotted lines correspond to WD~1 and WD~2, respectively, and the colours
		denote the combinations of BH spin, $\astar = \pm 0.98$ and orbit
		$\theta_a = 1^\circ, 90^\circ$ indicated in the legend.
		\textbf{Left Panel:} Binary with prograde rotation, $\vartheta_0 = 0^\circ$.
		\textbf{Right Panel:} Binary with retrograde rotation, $\vartheta_0 = 180^\circ$.}}	
	\label{rhomax1}
\end{figure}

\begin{figure}[h]
	\epsscale{1.1}
	\plottwo{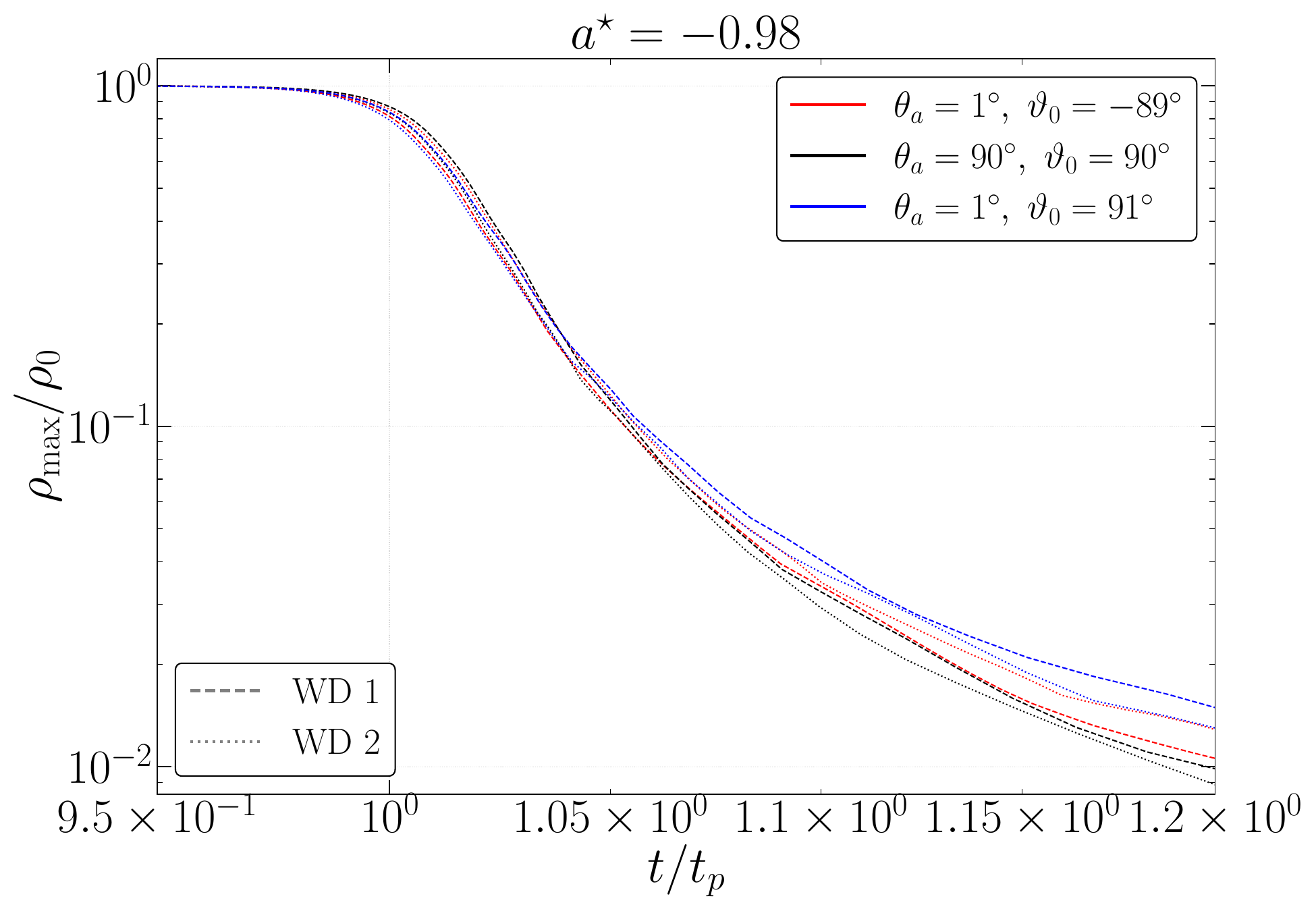}{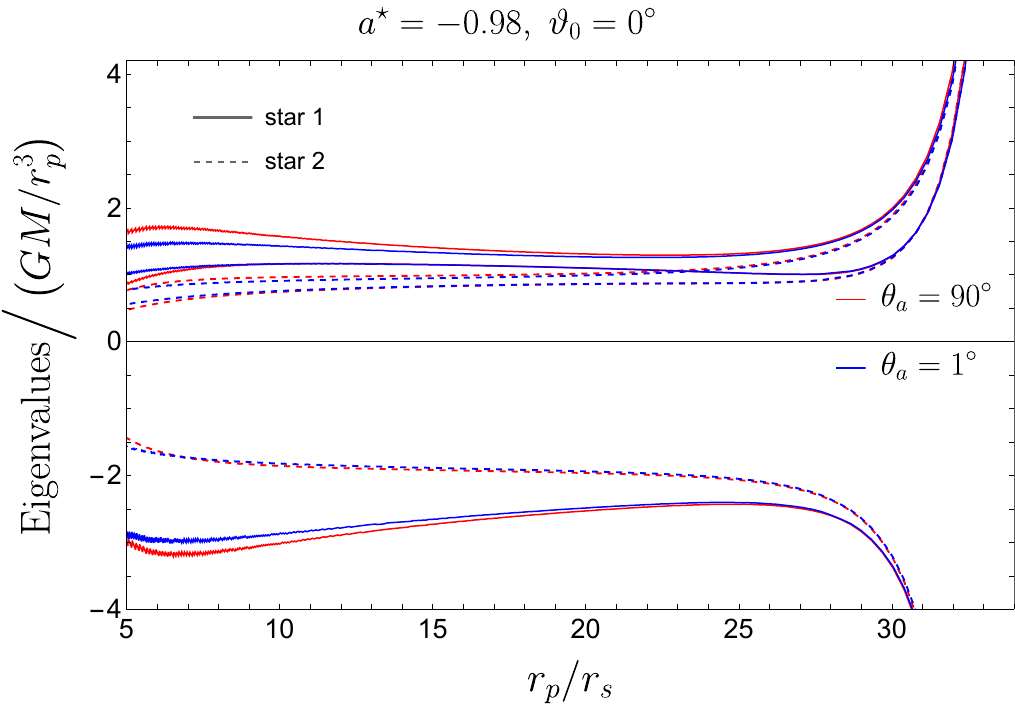}
	\caption{{\small \tb{\textbf{Left Panel:} Evolution of $\rho_{\max}/\rho_0$ as a function of
		$t/t_p$ for a BH with $\astar = -0.98$, with the binary spin perpendicular to the OAM.
		For the off-equatorial orbit ($\theta_a = 1^\circ$), the binary spin is along
		$+\hat{z}$ ($\vartheta_0 = -89^\circ$; red) or $-\hat{z}$
		($\vartheta_0 = 91^\circ$; blue), i.e., along the BH spin axis. For the
		equatorial orbit ($\theta_a = 90^\circ$, $\vartheta_0 = 90^\circ$; black), the OAM is
		along the BH spin axis, while the binary spin lies in the orbital plane, pointing
		towards the BH. Dashed and dotted lines correspond to WD~1 and WD~2, respectively.
		\textbf{Right Panel:} Eigenvalues of the tidal tensor $C_{ij}$, evaluated at pericentre
		and normalised by $GM/r_p^3$, as a function of the pericentre distance $r_p$ (in units
		of the Schwarzschild radius $r_s = 2r_g$), for a prograde binary
		($\vartheta_0 = 0^\circ$) around a BH with $\astar = -0.98$, for $\theta_a = 90^\circ$
		(red) and $\theta_a = 1^\circ$ (blue). The upper curves are the two positive
		(compressive) eigenvalues and the lower curves the negative (stretching) eigenvalue;
		solid and dashed lines correspond to star~1 and star~2, respectively.}}}	

	\label{rhomax2}
\end{figure}

To understand the relative tidal strength acting on each WD during the deformation
near pericentre, we first analyse the evolution of the maximum (central) density,
$\rho_{\rm max}/\rho_0$. Figure~\ref{rhomax1} shows the cases $\vartheta_0 = 0^\circ$ and
$\vartheta_0 = 180^\circ$ in the left and right panels, respectively. For
$\vartheta_0 = 0^\circ$, WD~1 is stretched more rapidly than WD~2, ultimately leading to
its complete disruption, and the differences among the orbital configurations are
consistent with the disruption behaviour shown in the left panel of Figure~\ref{mbound_pr}.
For $\vartheta_0 = 180^\circ$, the configurations differ slightly at early times, but
later the maximum densities increase again as the inherent interactions of the bound
binary begin to dominate. For $\vartheta_0 \simeq \pm 90^\circ$ (left panel of
Figure~\ref{rhomax2}), the differences are small.

\begin{figure}[h]
	\epsscale{1.1}
	\plottwo{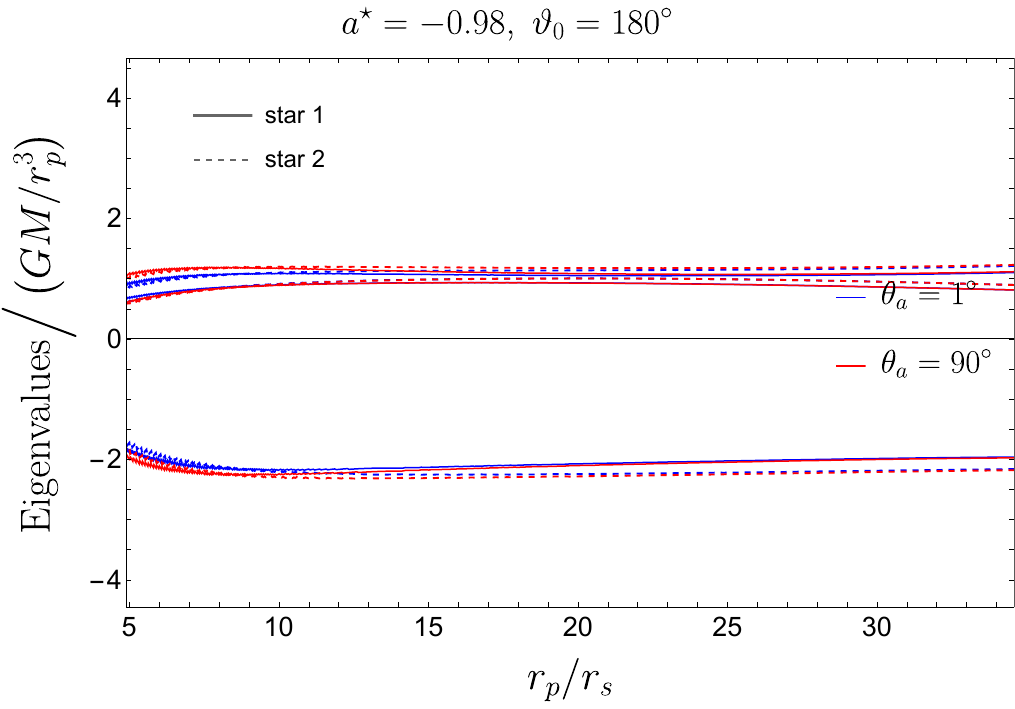}{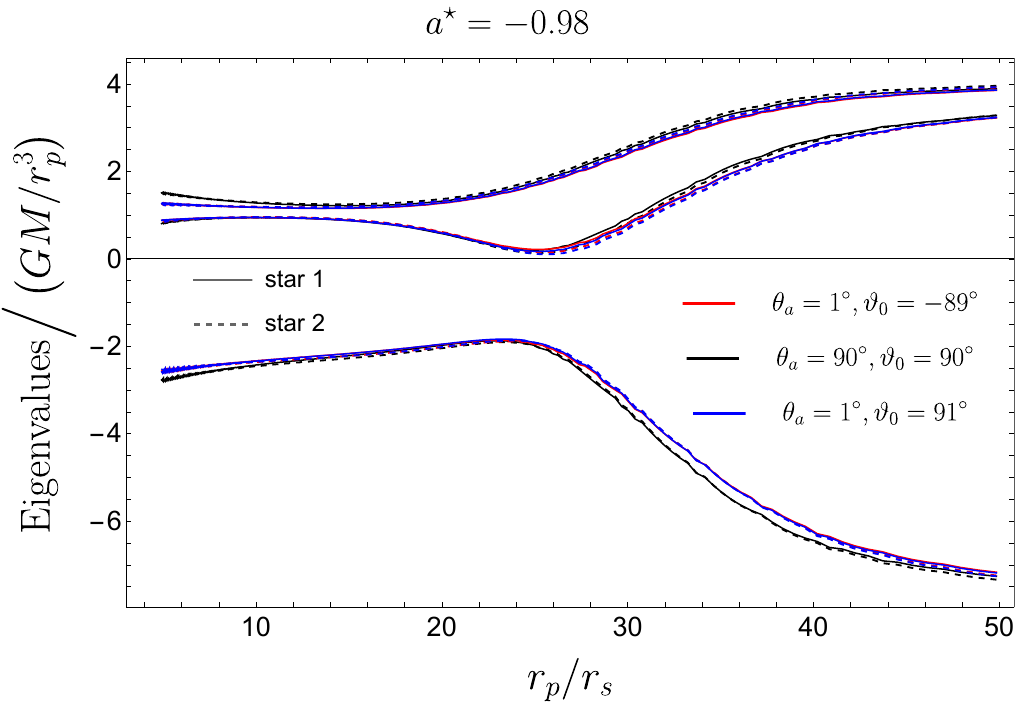}
	\caption{{\small \tb{Eigenvalues of the tidal tensor $C_{ij}$, evaluated at pericentre and normalised by $GM/r_p^3$, as a function of the pericentre distance $r_p$ (in units of $r_s = 2r_g$), for a BH with $a^\star = -0.98$. In both panels, the upper curves are the positive (compressive) eigenvalues and the lower curves are the negative (stretching)
		eigenvalue; solid and dashed lines correspond to star 1 and star 2, respectively. {\bf Left Panel:} Retrograde binary, $\vartheta_0 = 180^\circ$, for $\theta_a = 1^\circ$ (blue) and $\theta_a = 90^\circ$ (red).
		{\bf Right Panel:} Binary spin perpendicular to the OAM: $\theta_a = 1^\circ$ with $\vartheta_0 = 91^\circ$ (binary spin along $-\hat{z}$; blue) and $\vartheta_0 = -89^\circ$ (binary spin along $+\hat{z}$; red), and $\theta_a = 90^\circ$ with $\vartheta_0 = 90^\circ$ (binary spin in the orbital plane, pointing towards the BH; black).}}}
	\label{eigenvalue}
\end{figure}

To quantify the tidal field experienced by each star, we compute the eigenvalues of the
tidal stress tensor $C_{ij} = -\partial f_i/\partial x^j$ at pericentre, where the acceleration
$f_i = g_{i\nu}\ddot{x}^\nu$ is obtained from Equations \ref{KerrMetric} and \ref{eom}. We evaluate these
eigenvalues, in units of $GM/r_p^3$, over a range of pericentre distances, which also
shows how they depart from the Newtonian point-mass values $(1,1,-2)$. In each of
Figures~\ref{rhomax2} (right) and \ref{eigenvalue}, the upper curves denote the two positive
(compressive) eigenvalues and the lower curves the negative (stretching) eigenvalue. 
For $\vartheta_0 = 0^\circ$ (right panel of Figure~\ref{rhomax2}), at the pericentre
adopted in our simulations, $r_p \simeq 12.5\,r_s$ ($= 25\,r_g$), the eigenvalues show a
clear distinction between the two stars, with a smaller additional difference between
the $\theta_a = 1^\circ$ and $90^\circ$ configurations. These differences gradually
diminish with increasing $r_p$. Beyond $r_p \sim 30\,r_s$, however, the eigenvalues grow
rapidly in magnitude. We verified that, in this regime, the binaries undergo extreme stretching as their separation decreases and they collide near pericentre.
Since the BH tidal field is essentially Newtonian at these distances, with eigenvalues
close to $(1,1,-2)$ in units of $GM/r_p^3$, this growth arises from the tidal field of
the companion. For $\vartheta_0 = 180^\circ$ (left panel of Figure~\ref{eigenvalue}), the differences between
the orbital configurations are smaller than for $\vartheta_0 = 0^\circ$ and vanish at
larger $r_p$. \tb{For $\vartheta_0 \simeq \pm 90^\circ$ (right panel of Figure~\ref{eigenvalue}),
the differences between the configurations also remain small, and the curves fully
converge at $r_p \simeq 50\,r_s$. In this case, the two positive eigenvalues, which are nearly equal at small $r_p$, separate into two distinct branches at larger $r_p$.}

\tb{Thus, at $r_p \simeq 12.5\,r_s$, the eigenvalues nearly coincide for both stars and for
all orbital configurations when $\vartheta_0 = 180^\circ$ and $\vartheta_0 \simeq \pm
90^\circ$, whereas they differ noticeably for $\vartheta_0 = 0^\circ$. Consequently, for
$\vartheta_0 = 0^\circ$ the differences in the mass-disruption outcomes emerge
immediately after $t = t_p$. For the other inclinations, these differences become
significant only later, once the inherent interactions of the binary begin to dominate,
especially for $\vartheta_0 = 180^\circ$.}


\subsection{Fallback rates}

\begin{figure}[h]
	\epsscale{1.17}
	\plottwo{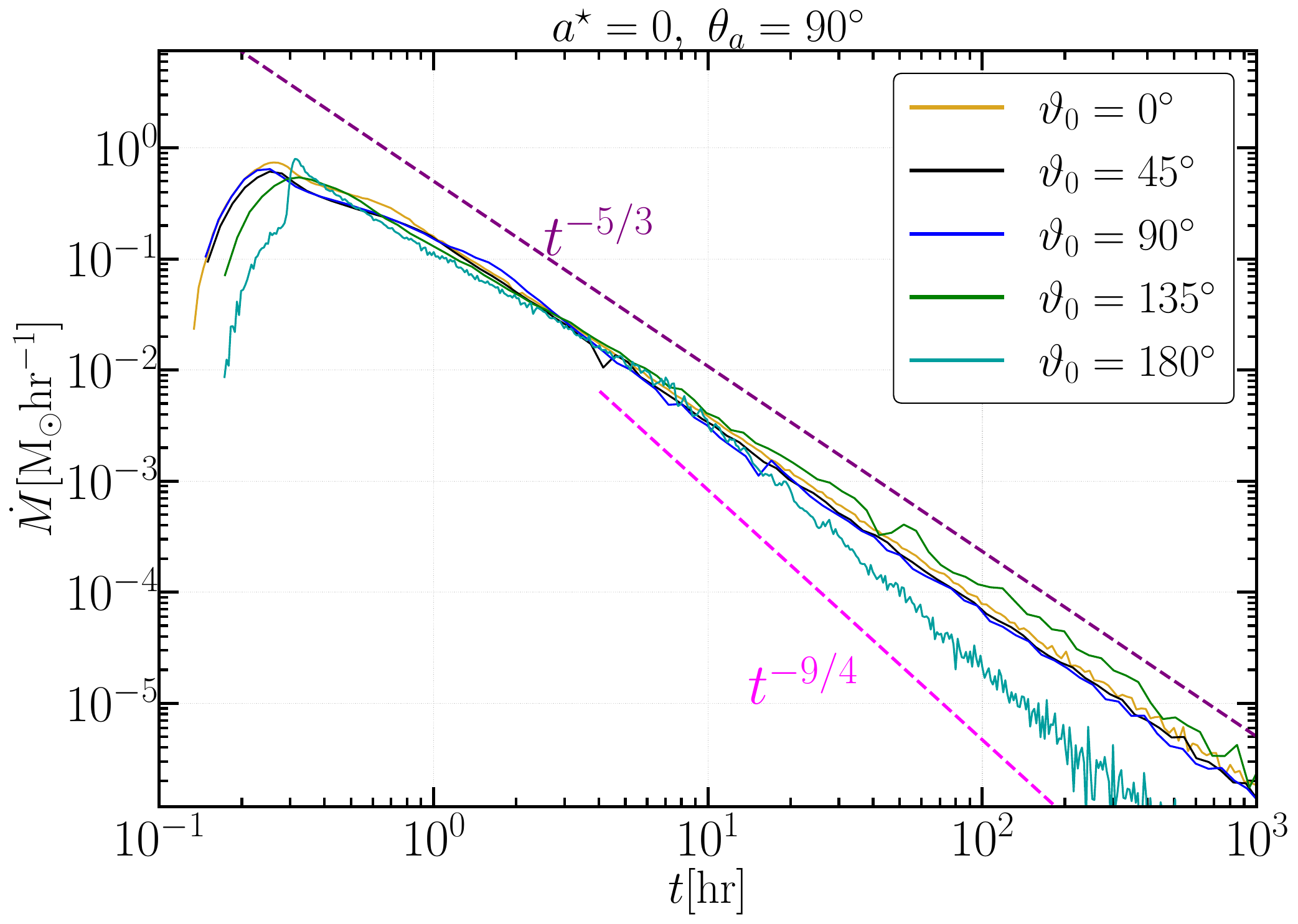}{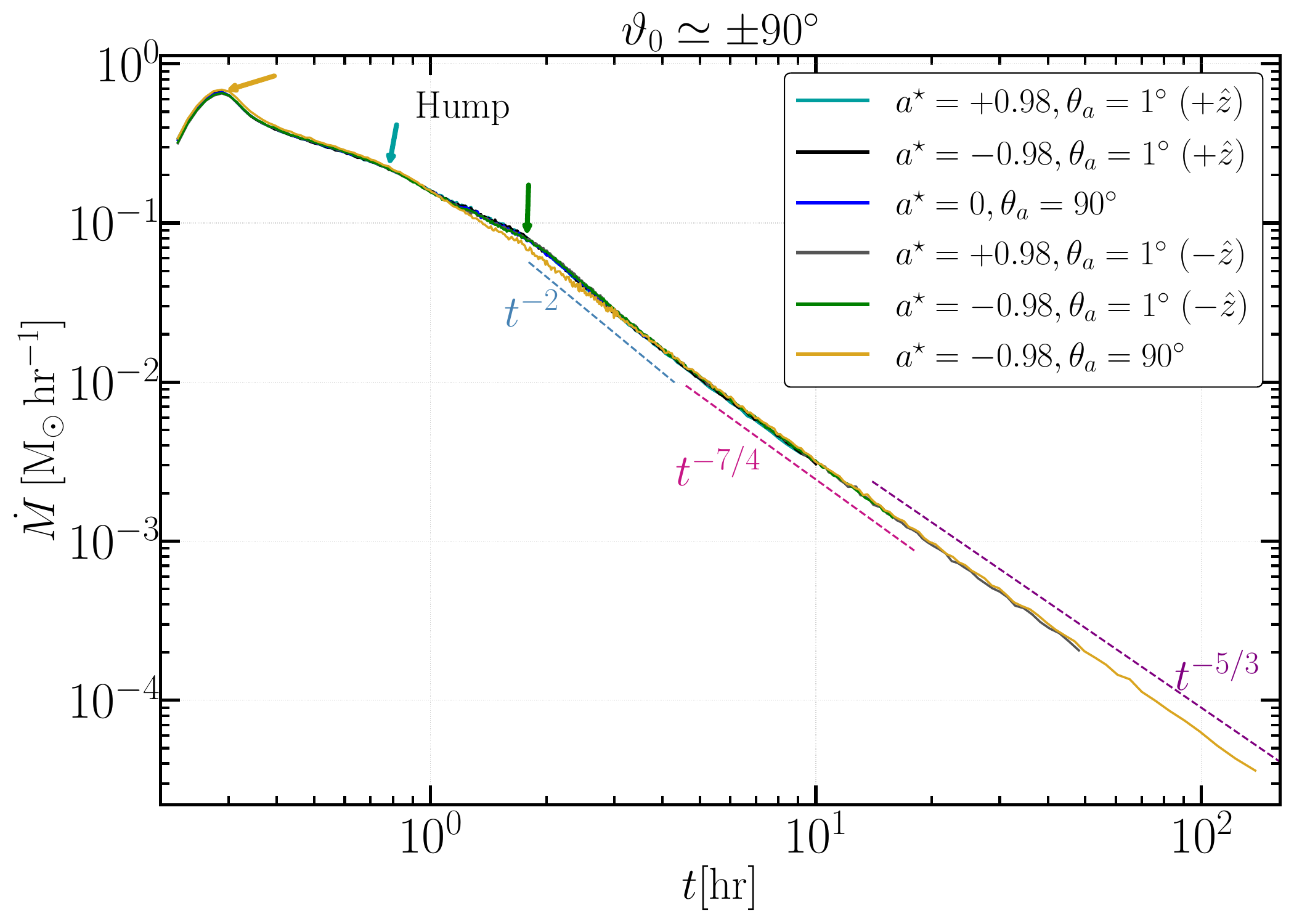}
	\caption{{\small Fallback rate, $\dot{M}$, as a function of time (in hours). 
			\textbf{Left panel:} Results for a non-spinning (Schwarzschild) BH for various initial binary inclinations $\vartheta_0$ and phase $\varphi_0=0^{\circ}$. 
			\textbf{Right panel:} Results for a spinning BH, with the initial binary spin aligned or anti-aligned with the $z$-axis ($\vartheta_0 \approx \pm 90^{\circ}$) for the $\theta_a = 1^{\circ}$ orbit. 
			For comparison, the cases \( a^{\star} = 0,\, \vartheta_0 = 90^{\circ} \) and \( a^{\star} = -0.98,\, \theta_a = -\vartheta_0 = 90^{\circ} \) are included, highlighting their similarity to the \( \theta_a = 1^{\circ} \) and $\vartheta_0 \approx \pm 90^{\circ}$. The arrows highlight the \emph{three-hump structure}, with the colours corresponding to different ranges of bound specific energy as discussed in the subsequent section.}}
	\label{dmdt_az}
\end{figure}

The left panel of Figure~\ref{dmdt_az} shows the fallback rates for the non-spinning black hole case (\(a^{\star}=0\), \(\theta_a = 90^{\circ}\)) for different binary inclinations \(\vartheta_0\), 
with the initial binary phase fixed at \(\varphi_0 = 0\). We find that the late-time fallback does not always follow the canonical \(t^{-5/3}\) power-law decay. In particular, the \(\vartheta_0 = 180^{\circ}\) configuration exhibits a more complex evolution: at intermediate times the fallback rate decays more slowly, approximately as \(t^{-7/5}\), before transitioning at later times to a steeper decline, close to \(t^{-9/4}\). In contrast, all other inclination angles adhere closely to the standard \(t^{-5/3}\) behaviour over the late-time regime. Further, as we predicted the nature of bound trajectories in section \ref{point_particle}, accordingly we observe that the peak fallback time, $t_{\rm peak}$ is obtained earlier when the disruption takes place in a more tightly bound orbit determined by the corresponding $\vartheta_0$. This diversity in fallback behaviour is notable and underscores the strong sensitivity of the fallback dynamics to the initial orientation of the binary. 

For the Schwarzschild case only, we compute the fallback rate using the frozen-in energy approximation. In all subsequent figures, the fallback rates are obtained by numerically differentiating the mass accumulated within a radius of \(3r_t\) as a function of time; nevertheless, the Schwarzschild result is retained in each panel for comparison. This provides a useful baseline against which the late-time behaviour of the spinning BH cases can be predicted.

\begin{figure}[h]
	\epsscale{1.17}
	\plottwo{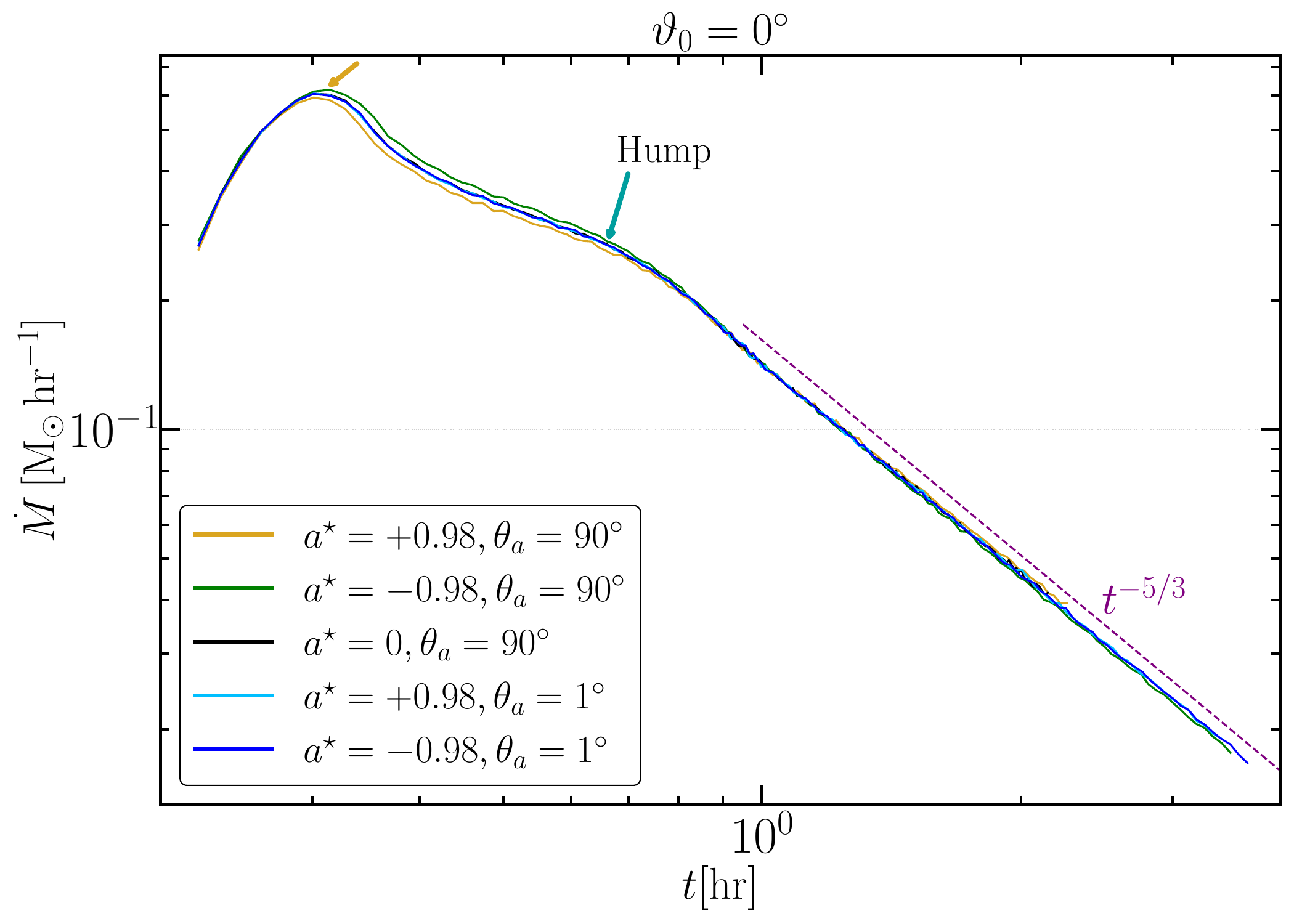}{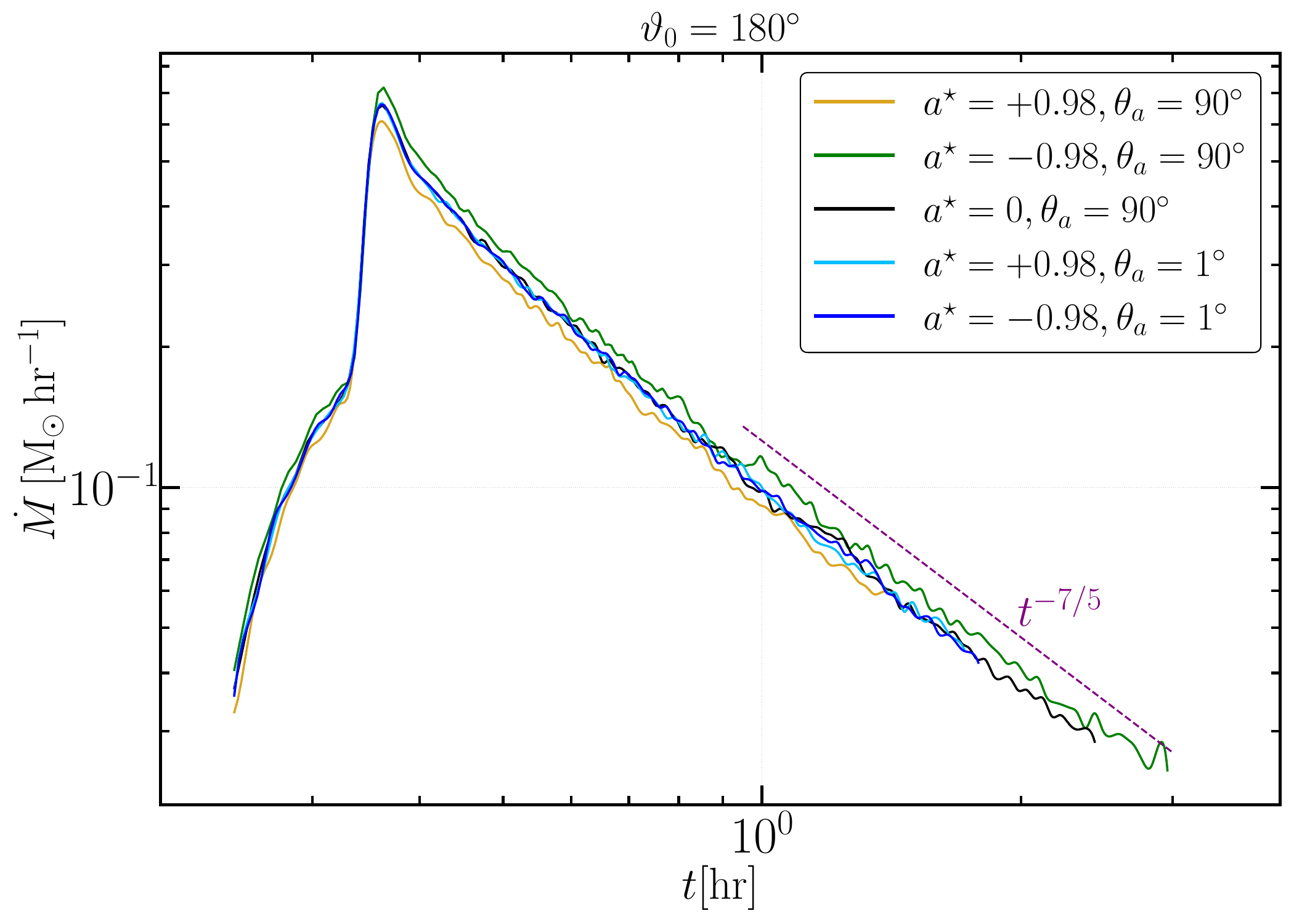}
	\caption{{\small Fallback rate, $\dot{M}$, as a function of time (in hours). 
			\textbf{Left panel:} Prograde binaries ($\vartheta_0=0^\circ$) orbiting a spinning BH, shown for both equatorial and off-equatorial orbits. The arrows highlight the double hump feature visible in the prograde case, with the colours corresponding to different ranges of bound specific energy as discussed in the subsequent section.
			\textbf{Right panel:} Retrograde binaries ($\vartheta_0=180^\circ$) orbiting a spinning BH, likewise shown for equatorial and off-equatorial orbits. 
			For comparison, the corresponding $a^{\star}=0$ (non-spinning BH) prograde and retrograde cases are included in each panel.}}
	\label{dmdt_pr}
\end{figure}

Introducing BH spin leads to distinct and systematic changes in the fallback behaviour. For prograde binaries (see left panel of Figure~\ref{dmdt_pr}), both in the equatorial and off-equatorial configurations, the fallback curves conform remarkably well to the $t^{-5/3}$ scaling across all orbital configurations. By contrast, retrograde binaries (see right panel of Figure~\ref{dmdt_pr}) deviate from this trend, displaying a shallower late-time decay approaching $t^{-7/5}$. The influence of BH spin in $\theta_a=90^{\circ}$ is also clearly reflected in the peak fallback rates, which follow the hierarchy 
\[
\dot{M}_{\rm peak}(a^{\star}=-0.98)
> \dot{M}_{\rm peak}(a^{\star}=0)
> \dot{M}_{\rm peak}(a^{\star}=+0.98).
\]

For retrograde encounters, the spin-induced differences persist even at late times: the $a^{\star}=-0.98$ case maintains the highest fallback rate among the three. This behaviour arises because the stronger apsidal precession associated with retrograde orbits brings the returning debris closer to the BH, enhancing the mass return rate.

An interesting behaviour emerges for binaries whose spin is aligned along $\pm z$ in $\theta_a=1^{\circ}$, as well as for systems with binary inclination $\vartheta_0 = 90^{\circ}$ in the equatorial plane (see right panel of Figure~\ref{dmdt_az}). In these configurations, the fallback curves display a distinct \emph{three-hump structure}, highlighted by arrows in the figure, that is different from the conventional two-hump pattern characteristic of typical binary disruptions. The late-time decay initially steepens to a slope close to \(t^{-2}\), before
transitioning to a shallower behaviour of approximately \(t^{-7/4}\). At even
later times, the fallback rate approaches the canonical \(t^{-5/3}\) scaling.
\tb{Such evolving power-law behaviour in time-dependent signals is also of interest in studies of turbulence and stochastic variability in strong-gravity environments \citep{Megale2025,Cook2026,Ficarra2026}.}
Comparison with the corresponding Schwarzschild case also indicates that, at
sufficiently late times, the fallback rate asymptotically recovers the standard
\(t^{-5/3}\) behaviour. This sequence of evolving power-law indices reflects a transient phase in which the debris dynamics are strongly influenced by relativistic effects and interactions, rather than a permanent departure from the standard fallback behaviour.

Note that stronger precession can drive debris closer to the black hole on subsequent passages, steepening the fallback rate at intermediate times, while the gradual phase mixing of the debris ultimately restores the decay expected from the frozen-in energy distribution.
Taken together, these results point to a richer and more intricate interplay between orbital geometry, internal binary orientation, and relativistic precession in tidal disruption events than previously appreciated.

\subsubsection{A brief note on the \textit{three-hump} structure}
\begin{figure}[h]
	\epsscale{0.9}
	\plottwo{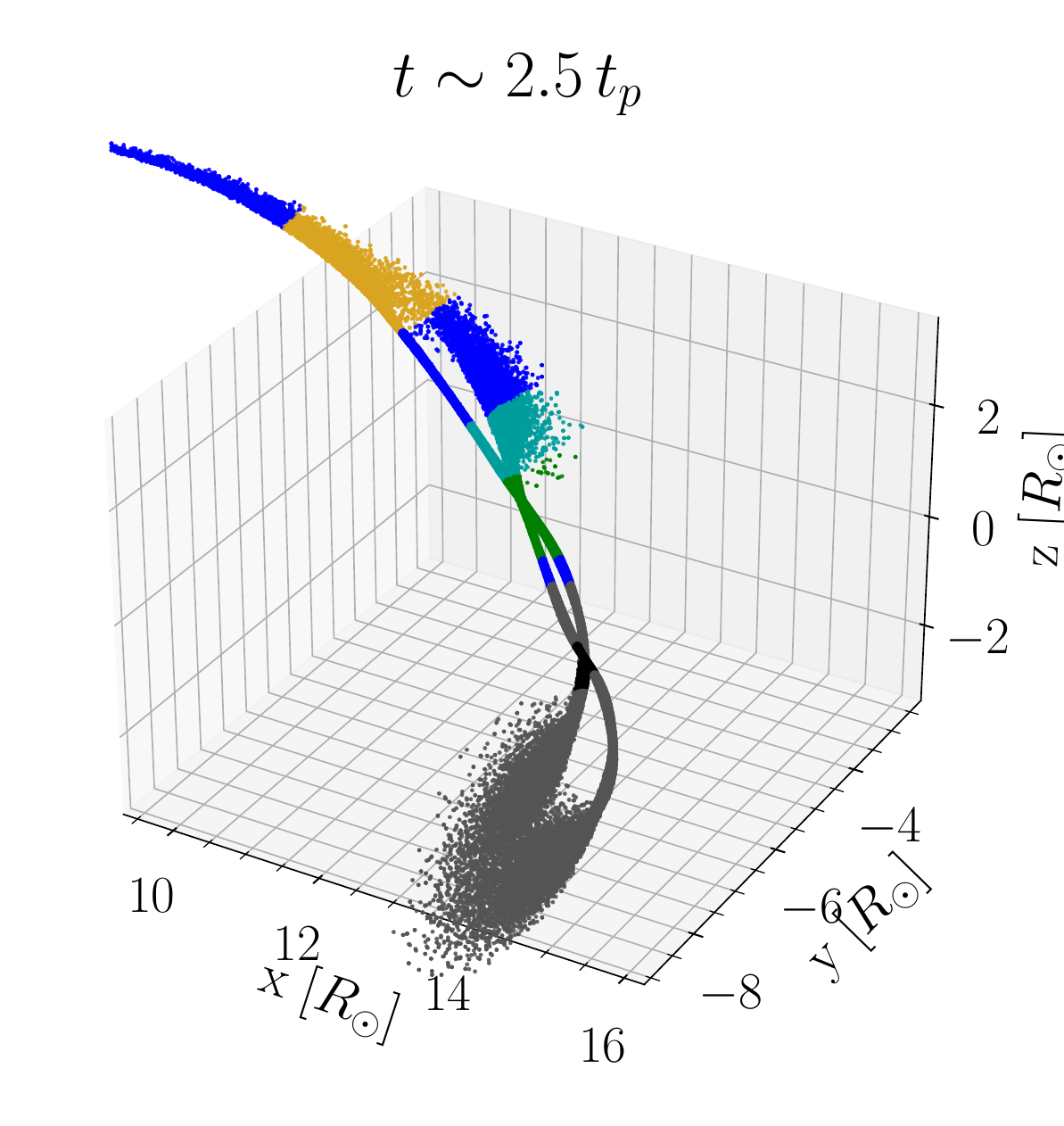}{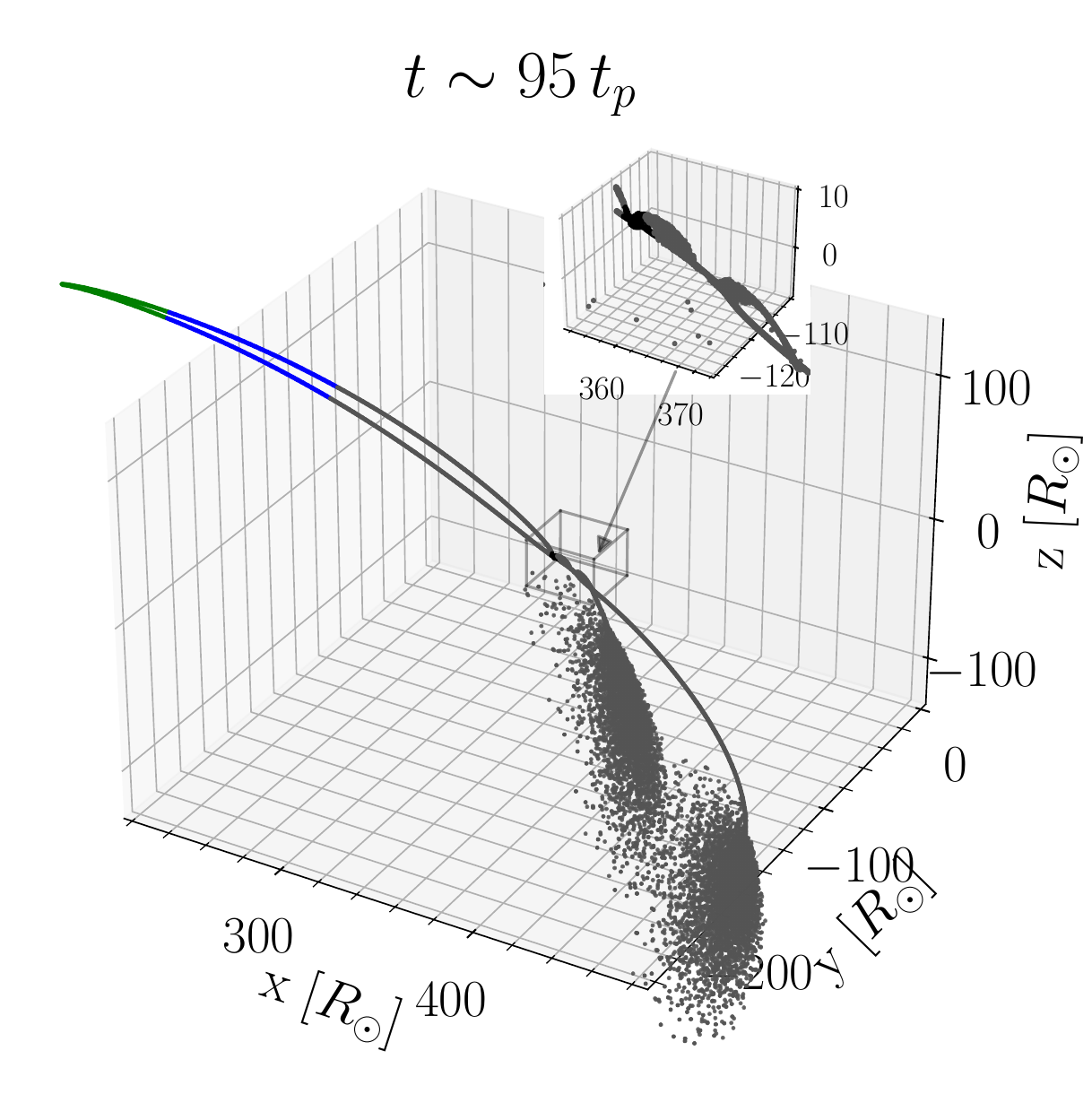}
	\caption{{\small Close-range snapshots from post-disruption simulations of a tidal encounter at inclination $\vartheta_0=90^\circ$ for $\astar=0$ case. The disrupted material is shown using a colour-coded distribution, where each colour corresponds to debris of specific energies. Yellow, cyan, green, and blue denote debris with negative specific energies, each contributing to a specific part of the fallback curve. While black marks the core particles and the rest of the material shown in grey has positive specific energy.			
			\textbf{Left panel:} Snapshot taken at $t\sim 2.5~t_p$, prior to the peak fallback rate $\dot{M}_{\rm peak}$. The tidal streams from both disrupted WDs wind around each other. The inner bound tails are accreting onto the BH, while the cores of both WDs are drawn together and undergo disruption. The unbound outer tails hangs behind the cores.
			\textbf{Right panel:} Snapshot taken at $t\sim 95~t_p$, where the late time behavior of fallback curve is guided by the remaining accreting debris (denoted by green and blue). The ejected WD has undergone partial disruption and its remnant core saturates at a lower mass. It holds the two outer tidal streams of both disrupted WDs, which are tied up together. An inset highlights this region with a clearer view.}}
	\label{snap}
\end{figure}

\begin{figure}[h]
	\epsscale{0.9}
	\plottwo{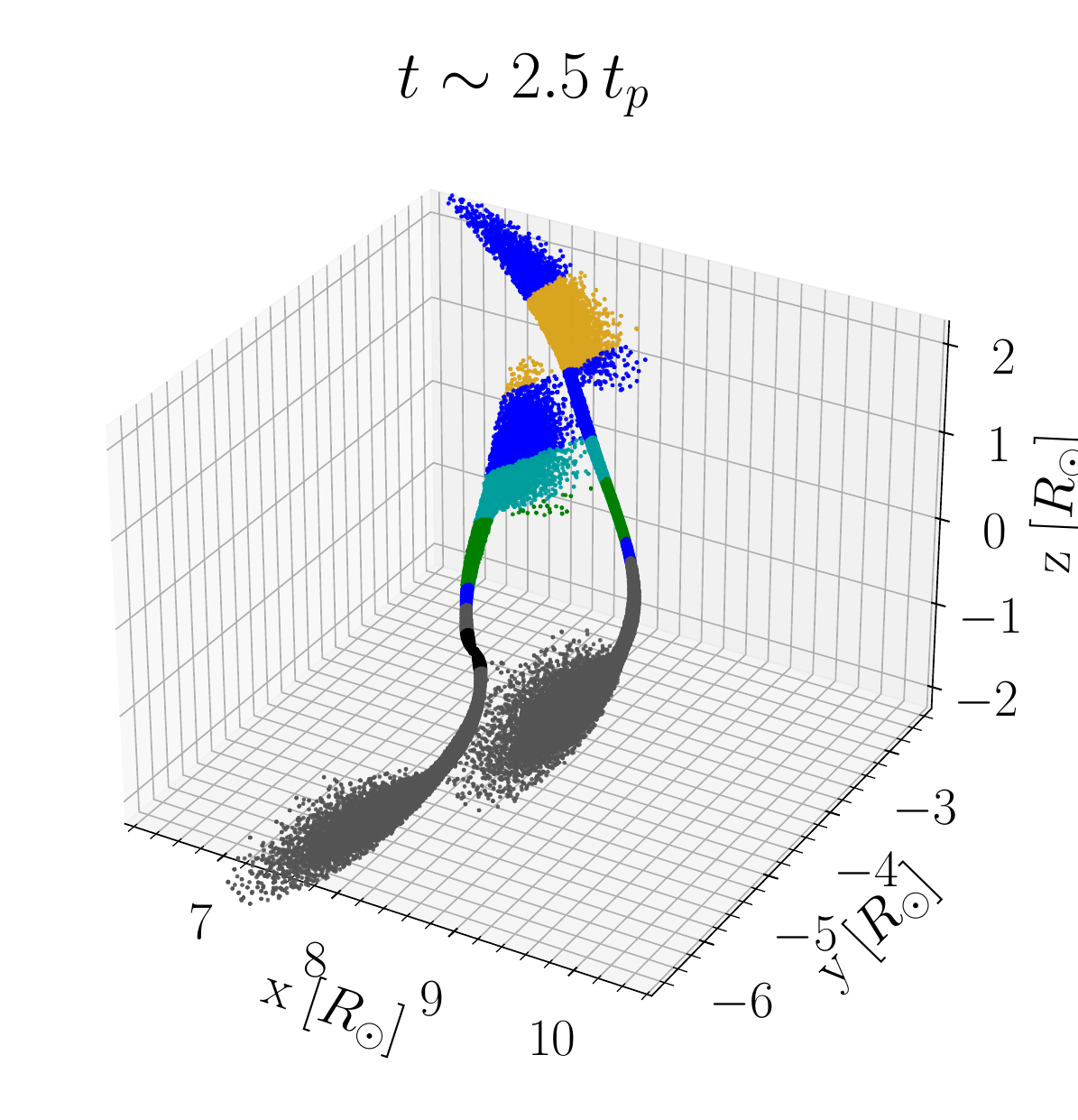}{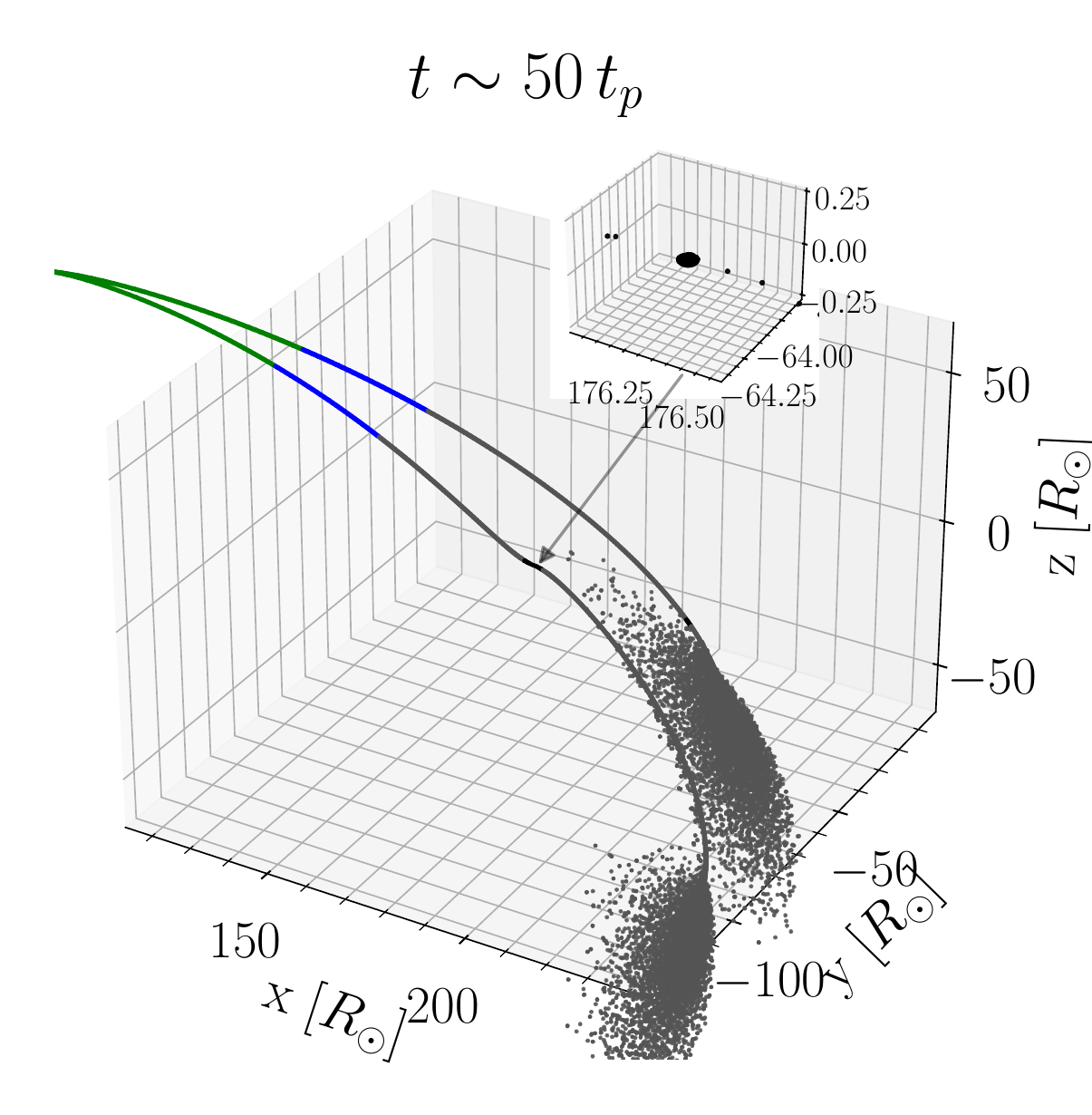}
	\caption{{\small Close-range snapshots from post-disruption simulations of a tidal encounter at inclination $\vartheta_0=0^\circ$ for $\astar=0$ case. \textbf{Left panel:} Snapshot taken at $t\sim 2.5~t_p$, prior to the peak fallback rate $\dot{M}_{\rm peak}$. The tidal streams from both disrupted WDs remains apart from each other.
			\textbf{Right panel:} Snapshot taken at $t\sim 50~t_p$. The ejected WD has undergone partial disruption and its remnant core saturates at a lower mass. An inset highlights this region with a clearer view.}}
	\label{snap2}
\end{figure}

In Figure \ref{snap}, we present the snapshots from two late-time stages of the post-disruption simulation for the $\astar=0$ and $\vartheta_0=90^\circ$ case. In the left panel, the inner bound tails accrete onto the BH, where the debris is 
shown in yellow, cyan, and green, each corresponding to a distinct range of negative specific energies. These contribute sequentially to the hump-like features seen in the fallback curve (see the right panel of Figure \ref{dmdt_az}). In contrast, for the $\vartheta_0 = 0^\circ$ configuration, Figure \ref{snap2} shows that only the bound debris (shown in yellow and cyan) contributes to the two-hump structure of the fallback curve (see the left panel of Figure \ref{dmdt_pr}). 

\begin{figure}[h]
	\epsscale{0.55}
	\plotone{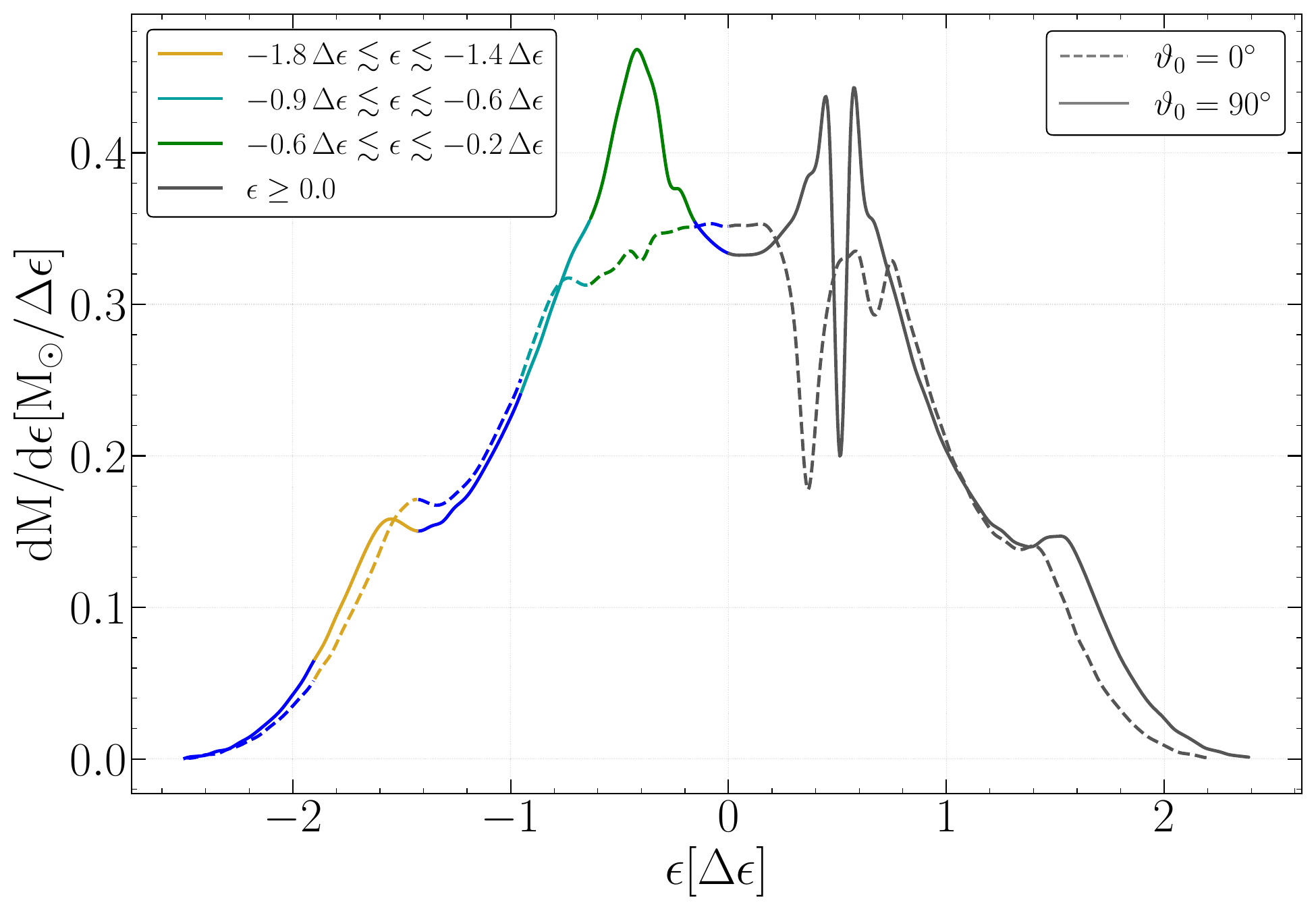}
	\caption{{\small The variation in the total debris differential mass distribution, $\rm{dM/d}\epsilon$ with respect to the specific energy $\epsilon$ for $\astar=0$ case. The solid and dashed lines denotes $\vartheta_0=0^\circ$ and $\vartheta_0=90^\circ$ configurations respectively. The coloured regions correspond to the indicated ranges of specific energies spanned by the debris. The remaining bound material outside these energy ranges is shown in blue.}}
	\label{dmde}
\end{figure}

We can comparatively estimate the overall distribution of the distrupted material by examining the total debris differential mass distribution, $\rm{dM/d}\epsilon$ with respect to specific energy in Figure \ref{dmde}. This distribution is quantified using the characteristic energy spread of the debris, $\Delta \epsilon = G M R_{\mathrm{bin}}/(r_t^b)^2$, where $R_{\mathrm{bin}} = R_{\mathrm{sep}} + R_{\mathrm{WD}}$. We find that the inner tail of WD1 containing a large fraction of the debris (shown in yellow), gives rise to the first hump in the fallback curve. Subsequently, the inner tail of WD2 together with a portion of the neck region of WD1 (collectively shown in cyan) contributes the second hump. Now, in the $\vartheta_0=90^\circ$ configuration, the third hump is produced by the merged region of the two tidal streams (shown in green), where maximum amount of debris are clustered around the same specific bound energy. This clustering of matter can enhance the role of self-gravity within that region, drawing additional material toward it and leading to more mass clumped within a narrow of range of specific energies, as seen in Figure \ref{dmde}. However, this is absent in $\vartheta_0=0^\circ$ configuration, where the debris in the same specific energy range (shown in green) from the two tidal streams remains separated from each other. As a result, the debris distribution in this range is nearly flat (see Figure \ref{dmde}), producing neither an additional hump nor a change in slope in the resulting fallback curve.

The remaining bound debris (shown in blue) contributes to the other portions of the fallback curve. Black marks the core particles and rest of the material shown in grey also do not contribute to the fallback curve, instead escape on a parabolic trajectory later.

%

\section{Discussions and conclusions}
\label{sec4}

In this paper, we have undertaken a comprehensive analysis of tidal interactions of a close WD binary with an IMBH. Apart from
the intrinsic interest in IMBHs for which observational literature is relatively less compared to their super-massive BH cousins, 
our results in this paper offer new insights into the rich physics of the dynamics of a WD binary in such BH backgrounds and highlights the 
intricate interplay between the BH spin, binary spin and binary OAM. We have considered cases when the CM of the binary is 
initially in the equatorial plane or close to the meridional plane, with an initial inclination between the binary OAM and spin 
to be $0^\circ, 90^\circ$ and $180^\circ$. These reasonably cover the possible range of the initial binary configuration, with 
maximally distinguishable results.

In particular, our results indicate that while BH spin non-trivially influences the outcome of the TDEs in terms of the core mass,
the binary inclination plays a central role in deciding the observables, for example the fallback rate of debris into the BH. 
In the previous work \cite{binary1}, we demonstrated how observable outcomes depend on the initial binary phase $\varphi_0$. From the fallback-rate calculations, we found that for prograde configurations the fallback curves were largely identical, whereas in retrograde configurations the detailed structure, such as the presence of humps, the late-time slope, and the peak fallback rate $\dot{M}_{\mathrm{peak}}$ varied significantly. This holds in a qualitatively similar way in the present study for $\vartheta_0=0^\circ$ and $\vartheta_0=180^\circ$. This happens because, for $\vartheta_0 = 180^\circ$, the binary tends to remain bound, leading to strong internal interactions between the disrupted companions, which in turn modify the fallback curve. In contrast, for $\vartheta_0 = 0^\circ$, the binary undergoes a clean Hills-type tidal separation, and the disrupted components do not interact, resulting in a similar fallback behaviour for both sets of debris. 
	
Similarly, for the $\vartheta_0 = 90^\circ$ configuration, we have checked from a few simulations where the tidal encounter again proceeds via a Hills-type mechanism even for different values of $\varphi_0$. As a result, the interactions between the disrupted components are relatively less (in some cases none), and the partially disrupted core is ejected along with the outer tidal tails. Consequently, the overall nature of the fallback curve is not strongly affected by variations in $\varphi_0$.
We further emphasize that the differences in tidal outcomes across orbital configurations (characterized by $\astar$ and $\theta_a$) and between the two WDs are maximum for the $\vartheta_0 = 0^\circ$ case. For other binary inclination angles, these differences become significant only after interactions between the disrupted companions begin to dominate, particularly in the $\vartheta_0 = 180^\circ$ configuration. Finally, we mention that the three-hump behaviour in the fallback rate that we have seen here is an artefact of binary spin and will also be present in non-spinning 
black hole backgrounds.

\begin{center}{\bf Acknowledgements}\end{center}

We acknowledge the support and resources provided by PARAM Sanganak under the National Supercomputing Mission, Government of India, 
at the Indian Institute of Technology Kanpur. The work of AM is supported by Prime Minister’s Research Fellowship by the Ministry of 
Education, Govt. of India. 

\begin{center}{\bf Data Availability Statement}\end{center}

\tb{Owing to the large size of the simulation outputs, the data underlying this article will be shared upon reasonable request to the corresponding author. Upon receiving a request, the requested data (processed data underlying the figures and/or raw simulation snapshots) will be deposited in a public repository such as Zenodo, and the link will be shared along with documentation of the data format and scripts for reading and analysing the data.}

\appendix 
\section{Post Newtonian self-gravity}\label{appendix1}

\tb{We use the expression for the coordinate acceleration of body $a$ in the $\mathtt{1PN}$ dynamics of an isolated system of gravitating bodies, as derived in \cite{Will:2016sgx}. We consider the leading post-Newtonian correction in the following expansion,
\begin{equation}
	 \mathtt{a}_a^i =  \mathtt{a}_a^i [0\mathtt{PN}] + \mathtt{a}_a^i [1\mathtt{PN}] + \mathcal{O}(c^{-3})
\end{equation}}

\tb{For our binary components $m_a$ and $m_b$, we include their mutual gravitational interaction through $1\mathtt{PN}$ order, i.e. corrections of order $\mathcal{O}(c^{-2})$, in the self-gravity term of Equation \ref{eom}.}

\tb{\begin{eqnarray}
	(\ddot{x}_a^i)_{\rm sg} =&& -\frac{G m_b}{ |x_a^k - x_b^k|^{3}} (x_a^i-x_b^i) \left[1 - \frac{4 G m_b}{c^2 |x_a^k - x_b^k|} - \frac{5 G m_a}{c^2 |x_a^k - x_b^k| } + \frac{\dot{x}_a^k \dot{x}_{a,k}}{c^2} - \frac{4\dot{x}_a^k.\dot{x}_{b,k}}{c^2} + \frac{2\dot{x}_b^k \dot{x}_{b,k}}{c^2} - \frac{3}{2c^2} \left(\frac{\dot{x}_{b,j} (x_a^j - x_b^j)}{|x_a^k - x_b^k|}\right)^2 \right]\nonumber \\
&&+\frac{G m_b}{c^2 |x_a^k - x_b^k|^{3}}(x_a^j - x_b^j)(4 \dot{x}_{a,j} - 3\dot{x}_{b,j}) (\dot{x}_a^i - \dot{x}_b^i)
	\end{eqnarray}}
	
\tb{The $1\mathtt{PN}$ correction is applied only to the mutual gravitational interaction of the binary components, while the motion in the external Kerr spacetime is retained through the first term of Equation \ref{eom}. Thus, the equation of motion becomes
\begin{eqnarray}\label{eom2}
\ddot{x}^i_a = - \left( g^{i\lambda} - g^{0\lambda} \dot{x}^i \right)_a \left[ \left( \frac{\partial g_{\mu\lambda}}{\partial x^\nu} - \frac{1}{2} \frac{\partial g_{\mu\nu}}{\partial x^{\lambda}} \right) \dot{x}^{\mu} \dot{x}^{\nu} \right]_a + (\ddot{x}_a^i)_{\rm sg} 
\end{eqnarray}}

\tb{We solve Equation \ref{eom2} and compare the test-particle trajectories obtained using Newtonian and $1\mathtt{PN}$-corrected mutual gravitational interactions. We consider two distinct cases: a prograde rotation, in which the binary eventually becomes separated, and a retrograde rotation, in which the binary remains bound throughout the evolution.
\begin{figure}[h]
	\epsscale{1.0}
	\plottwo{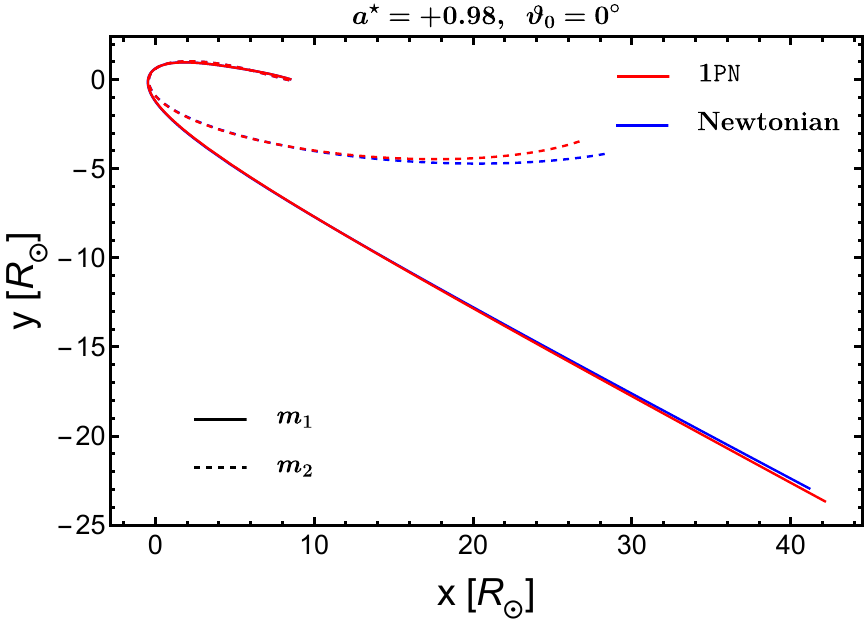}{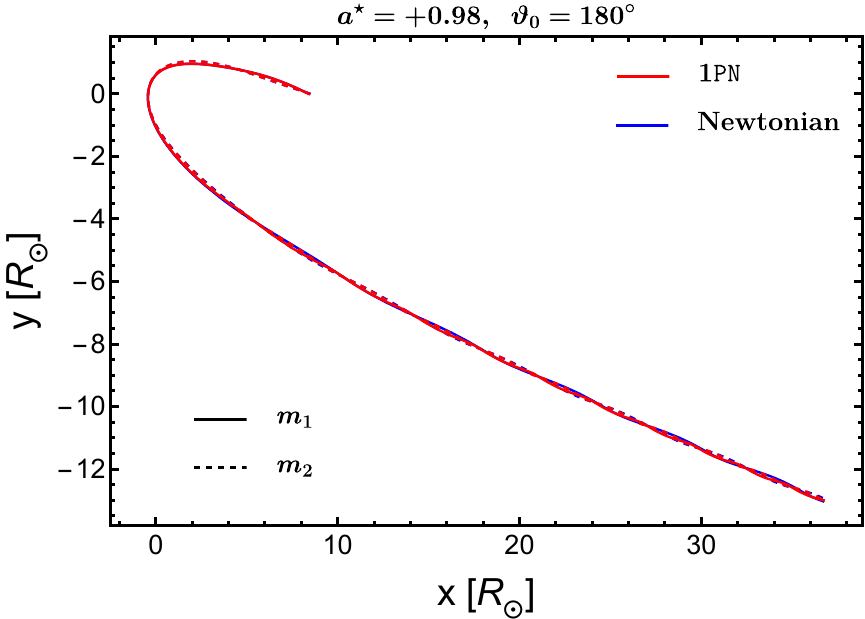}
	\caption{{\small Comparison of trajectories of binary test particle bound by Newtonian and $1\mathtt{PN}$ corrected self gravity around the BH with spin $\astar=+0.98$ on the equatorial orbit. \textbf{Left Panel: }The initial motion of the binary is prograde relative to the OAM ($\vartheta_0=0^\circ$). \textbf{Right Panel: }The initial motion of the binary is retrograde relative to the OAM ($\vartheta_0=180^\circ$).}}
	\label{xyz3}
\end{figure}}

\tb{In the left panel of Figure \ref{xyz3}, corresponding to the prograde rotation, the trajectories obtained with Newtonian and $1\mathtt{PN}$-corrected mutual gravity remain nearly identical during the initial evolution, but a relative deviation becomes noticeable as the binary separation increases at later times. This deviation results from the accumulation of the relativistic correction when the two binary components become more widely separated. In contrast, in the retrograde rotation shown in the right panel, the binary remains tightly bound and its separation stays relatively small throughout the evolution. Consequently, the $1\mathtt{PN}$ correction remains negligible, and the Newtonian and $1\mathtt{PN}$ trajectories show almost no discernible relative deviation.}

\section{Non-relativistic regime of the SPH particles}
\label{appendix:SPH_velocity}

\begin{figure}[h]
	\epsscale{1.0}
	\plottwo{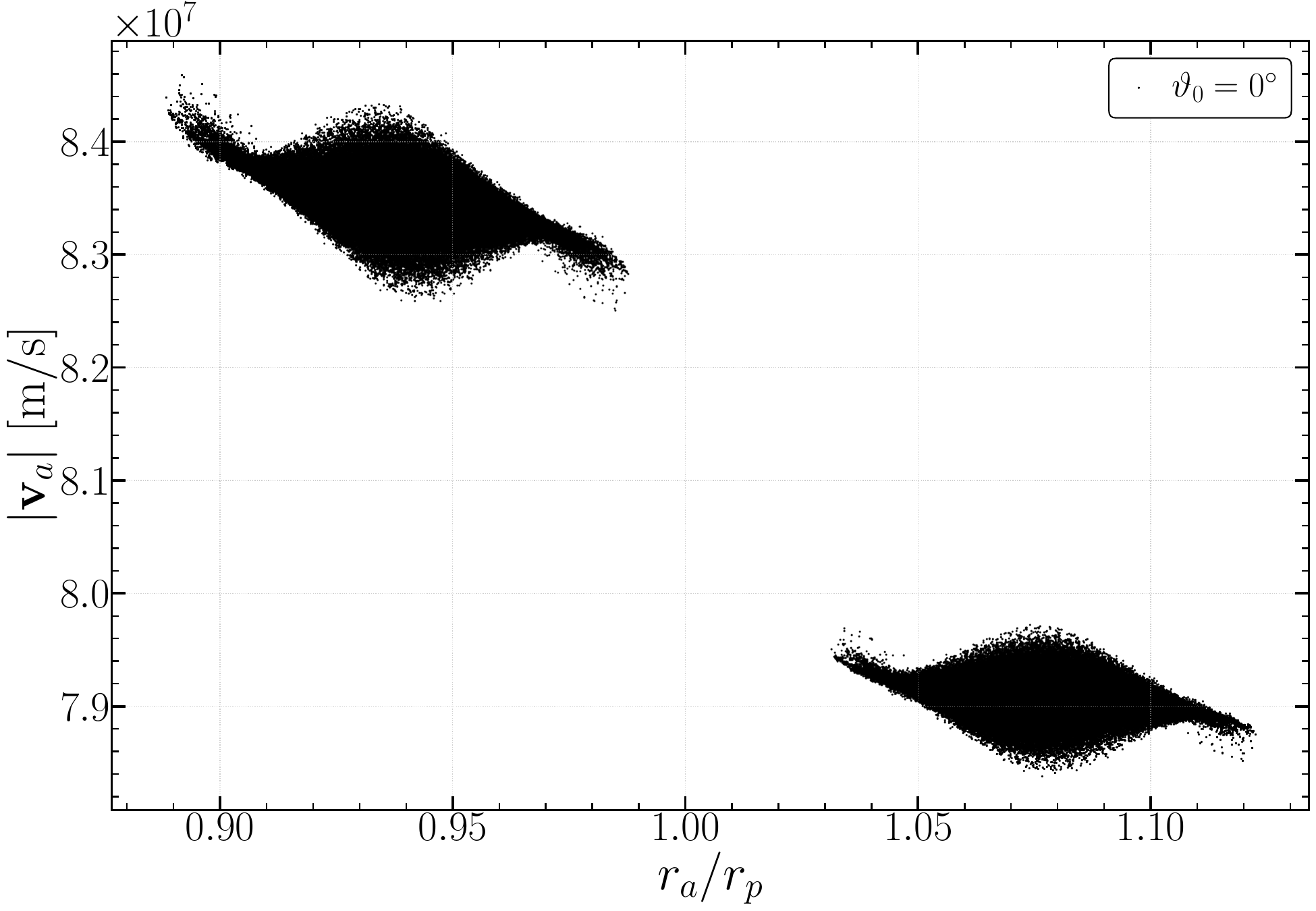}{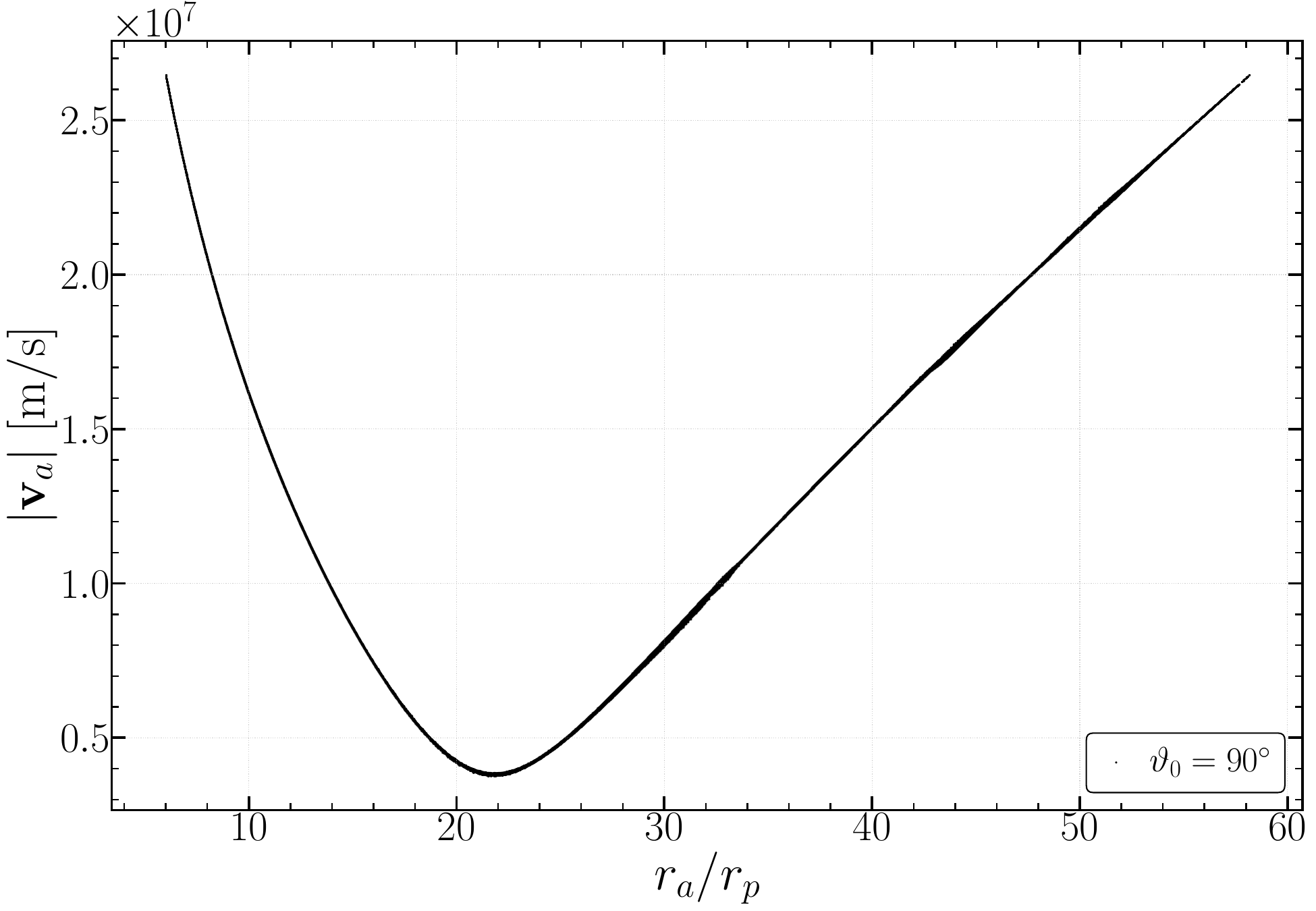}
	\caption{{\small Velocities of SPH particles presented for BH with spin $\astar=0.0$ corresponding to two different cases: \textbf{Left Panel: } $\vartheta_0=0^\circ$, at the pericentre, $t=t_p$, where $t_p\approx0.062~\mathrm{hr}$. \textbf{Right Panel: }$\vartheta_0=90^\circ$, after the pericentre passage, at $t=2.5~t_p$, when the tidal streams intersect around each other (as displayed in the left panel of Figure \ref{snap2}).}}
	\label{vel}
\end{figure}

\tb{To further verify the validity of the Newtonian hydrodynamic treatment, we examine the velocities of all SPH particles in two different cases in Figure \ref{vel} and at different dynamical times. The Lorentz factor of an SPH particle $a$ is given by $	\gamma_a=1/\sqrt{1-v_a^2/c^2}$, where $v_a=|\mathbf{v}_a|$ is the magnitude of the velcity of SPH particle $a$.}

\tb{The left panel of Figure \ref{vel} corresponds to the prograde configuration ($\vartheta_0=0^\circ$) at pericentre, where the SPH particles attain their maximum velocities in our simulations. The maximum velocity is approximately
$v_a^{\max}\simeq8.46\times10^7~{\rm m\,s^{-1}}\simeq0.282c$,
corresponding to a maximum Lorentz factor
$\gamma_a^{\max}\simeq1.042$. The right panel corresponds to a different inclination, $\vartheta_0=90^\circ$, at $t=2.5~t_p$, after the pericentre passage, when the tidal debris streams intersect and can potentially generate tidal shocks. Here, the maximum SPH-particle velocity is approximately
$v_a^{\max}\simeq2.7\times10^7~{\rm m\,s^{-1}}\simeq0.090c$,
giving $\gamma_a^{\max}\simeq1.004$. These maximum velocities are also seen in other configurations (including $a^\star=\pm0.98$) in the same order of magnitude.}

\tb{Thus, even at the pericentre passage, where the SPH particles attain their largest velocities, the Lorentz factor remains close to unity. Moreover, during the subsequent intersection of the tidal debris, where shock heating and dissipative hydrodynamic processes can become important, the particle velocities remain well within the non-relativistic regime. Therefore, relativistic corrections to the particle kinematics are small, and the hydrodynamic processes, including the associated artificial-viscosity dissipation used to capture shocks, can be consistently treated within the Newtonian framework.}

\section{Resolution Dependence}\label{appendix:res}

\begin{figure}[h]
	\epsscale{1.0}
	\plottwo{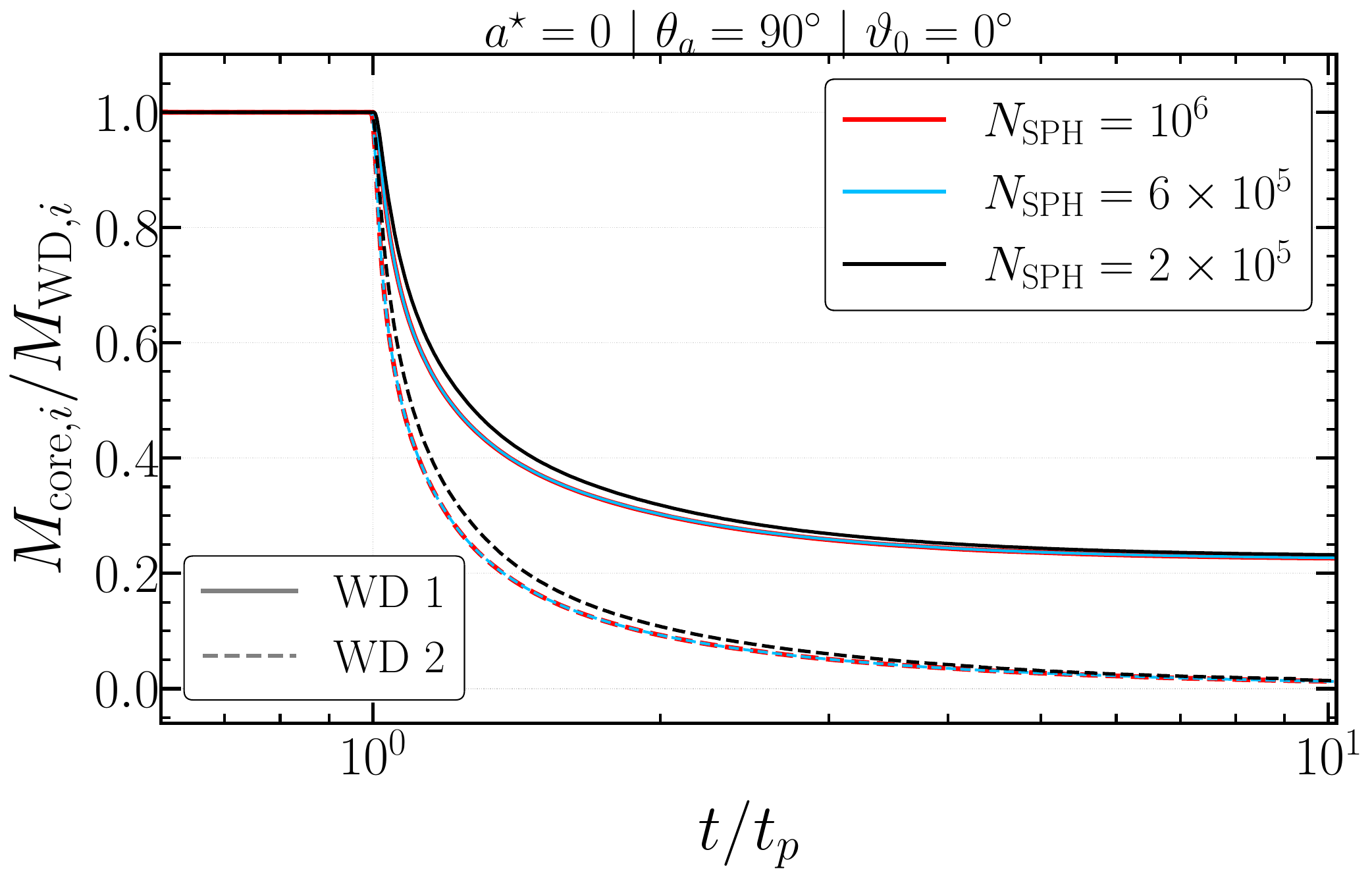}{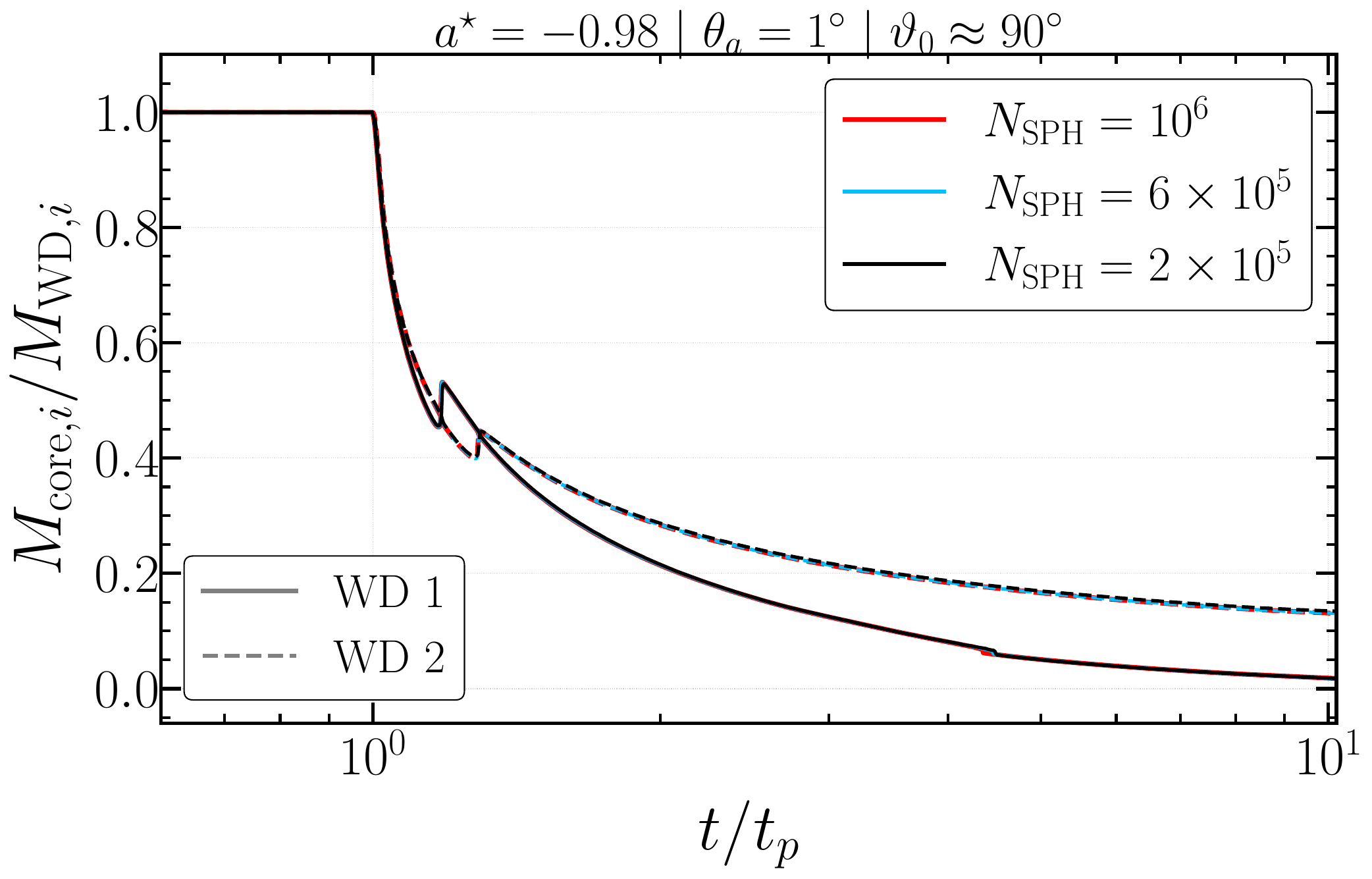}
	\caption{{\small Resolution dependence of the core mass fraction $M_{\rm core}/M_{{\rm WD},i}$. The red, skyblue, and black curves correspond to
			$N_{\rm SPH} =10^6$, $N_{\rm SPH} =6 \times 10^5$, and 	$N_{\rm SPH} =2 \times 10^5$, respectively, and are visually overlapping, indicating good numerical convergence. \textbf{Left Panel: } Binary in prograde rotation encountering a non-spinning BH on an equatorial orbit. \textbf{Right Panel: }Binary with initial inclination $\vartheta_0\approx90^\circ$ encountering spinning BH with $a^\star=-0.98$ on an off-equatorial orbit with $\theta_a=1^\circ$.}}
	\label{res}
\end{figure}

\begin{figure}[h]
	\epsscale{1.0}
	\plottwo{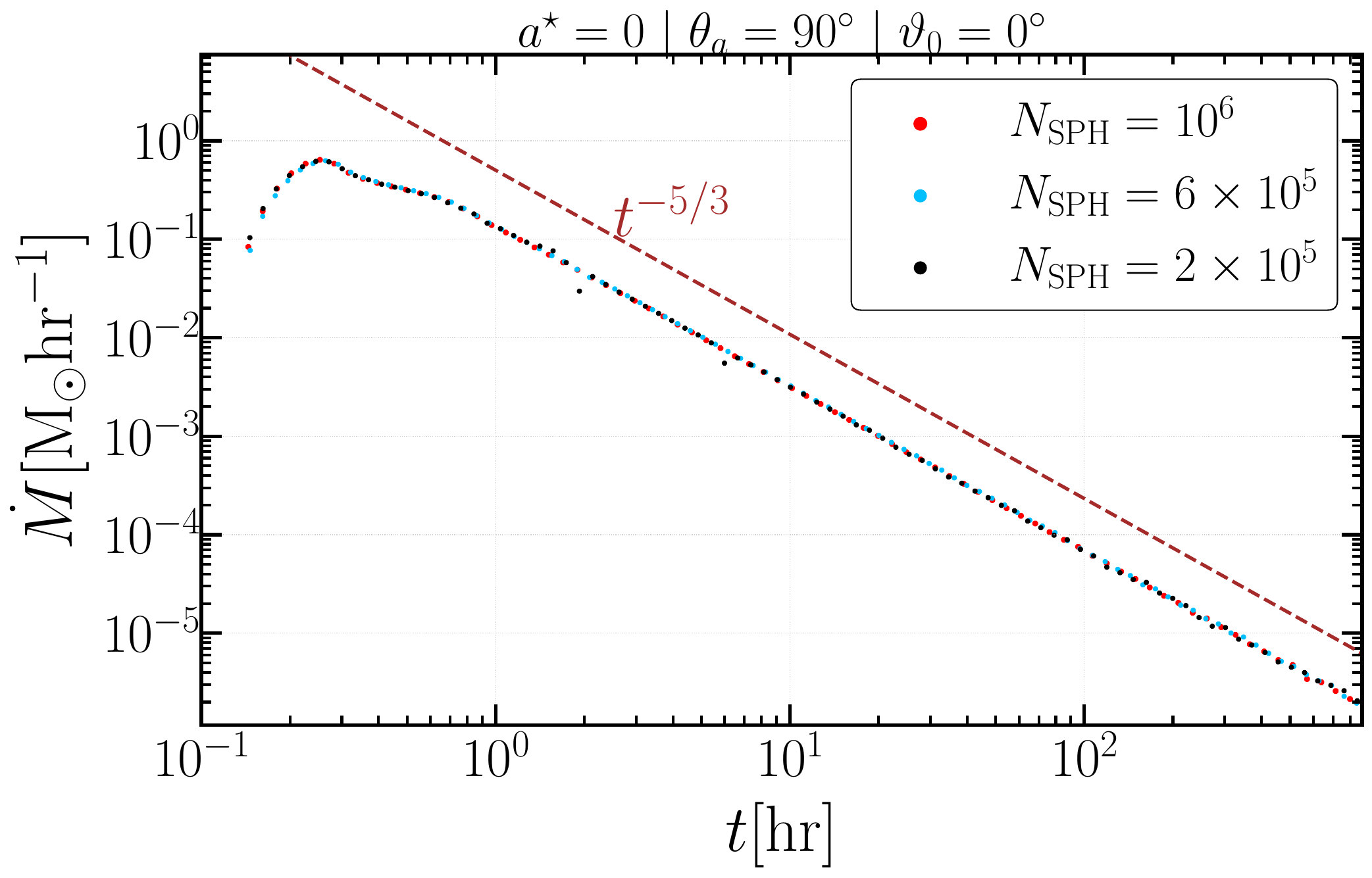}{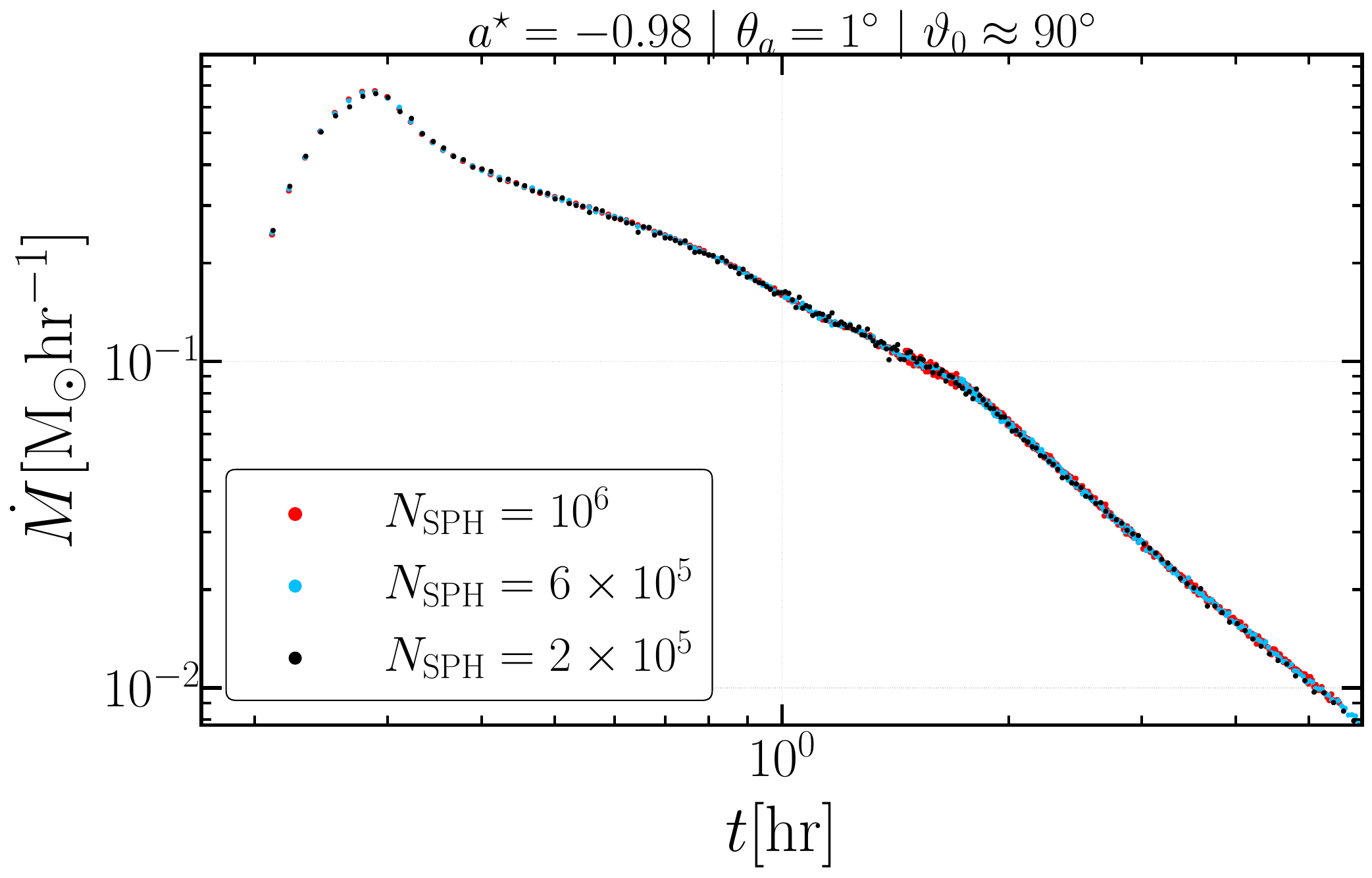}
	\caption{{\small Resolution dependence of the fallback rates $\dot{M}$. The red, skyblue, and black curves correspond to
		$N_{\rm SPH} =10^6$, $N_{\rm SPH} =6 \times 10^5$, and 	$N_{\rm SPH} =2 \times 10^5$, respectively, and are visually overlapping, indicating good numerical convergence. \textbf{Left Panel: } Binary in prograde rotation encountering a non-spinning BH on an equatorial orbit. \textbf{Right Panel: }Binary with initial inclination $\vartheta_0\approx90^\circ$ encountering spinning BH with $a^\star=-0.98$ on an off-equatorial orbit with $\theta_a=1^\circ$.}}
	\label{res2}
\end{figure}

\tb{To verify the convergence of our simulation results with respect to SPH particle
resolution, we compare the outcomes obtained with $N_{\rm SPH}=2\times10^5$
and $10^6$ particles (or $10^5$ and $5\times 10^5$ particles per WD) with those of our present resolution,
$N_{\rm SPH}=6\times10^5$ (or $3\times 10^5$ particles per WD). We consider two representative cases; a
non-spinning BH encountered by a prograde binary, and a spinning BH
encountered by an inclined binary on an off-equatorial orbit. The resolution
dependence of the core mass fraction and the fallback rate is shown in
Figures~\ref{res} and~\ref{res2}, respectively.}

\tb{For the core mass fraction, $M_{\rm core}/M_{{\rm WD},i}$, the lowest-resolution
run exhibits a small initial deviation from the higher-resolution results.
However, the saturated core masses are well converged in both panels. The
maximum relative deviations from the $N_{\rm SPH}=10^6$ results are
$\lesssim0.85\%$ for $N_{\rm SPH}=2\times10^5$ and
$\lesssim0.21\%$ for $N_{\rm SPH}=6\times10^5$.}

\tb{The fallback rates in Figure~\ref{res2} likewise show good convergence across
the three resolutions. In particular, the characteristic features of the
fallback curves are preserved, including the double-hump structure and the
late-time $t^{-5/3}$ decline in the left panel, as well as the triple-hump
structure in the right panel. The maximum relative deviations in the peak
fallback rate, $\dot{M}_{\rm peak}$, are also below $1\%$. These comparisons
demonstrate that the principal features of both variables and
observables calculated in our study are robust against the choice of SPH
resolution.}

\end{document}